\documentclass[12pt,a4paper]{article}
\usepackage{a4wide}
\usepackage{amsmath}
\usepackage{amssymb}
\usepackage{epsfig}
\usepackage{subfigure}
\usepackage{exscale}
\usepackage{float}
\usepackage{bbm}
\usepackage[numbers,sort&compress]{natbib}
\usepackage{pst-plot, pstricks,pst-math}
\usepackage{fancybox,amssymb,color}
\usepackage{graphicx}
\usepackage{pstricks, color, graphicx, epsfig, psfrag}
\usepackage{amsfonts,amsmath,amssymb,slashed}
\usepackage{dsfont}
\usepackage{bbm,bm}
\usepackage{fancyhdr, a4wide}
\usepackage[english]{babel}
\usepackage{subfigure}

\makeatletter
\@addtoreset{equation}{section}
\makeatother

\title{Thermodynamics of the $d$=3+1 Quantum XY Model}

\author{Christoph P.\ Hofmann$^a$ \\ \\
\normalsize{$^a$ Facultad de Ciencias, Universidad de Colima} \\
\vspace{0.3cm}
\normalsize {Bernal D\'iaz del Castillo 340, Colima C.P.\ 28045, Mexico} \\}

\begin{document}

\maketitle

\begin{abstract} \normalsize

Within effective field theory we explore the properties of the $d$=3+1 quantum XY model at low temperatures and in weak magnetic or
staggered fields. For this parameter regime only few results appear to be known, and furthermore are restricted to one-loop order. In the
present study we systematically analyze the thermodynamics of the $d$=3+1 quantum XY model up to three-loop order. In the low-temperature
expansion of the free energy density, the free Bose gas term of order $T^4$ receives corrections of order $T^6$ and $T^8$. The discussion
also includes the pressure, (staggered) magnetization and susceptibility. In particular, we show how these quantities are influenced by
the spin-wave interaction. We then compare our findings with those for the quantum XY model in $d$=2+1.

\end{abstract}


\maketitle

\section{Introduction}
\label{Intro}

Whereas the finite-temperature properties of the $d$=2+1 quantum XY model have received a lot of attention over the past few decades --
presumably due to the occurrence of the Kosterlitz-Thouless phase transition -- the same can not be said about the quantum XY model in 
$d$=3+1. Apart from some pioneering articles, no systematic studies of its thermodynamic properties at low temperatures, where the
spin-wave picture applies, seem to be available. The present work closes this gap that apparently exists in the condensed matter
literature.

As is well-known, the relevant low-energy excitations of the $d$=3+1 quantum XY model are the spin waves. They emerge as a consequence of
the spontaneously broken internal symmetry O(2) $\to$ 1. The low-energy behavior of the system can thus be captured by the physics of its
Goldstone bosons. This is the effective-field theory point of view we pursue in the present study. In contrast to spin-wave theory or
other microscopic approaches, the effective Lagrangian method allows one to study the low-temperature behavior of the $d$=3+1 quantum XY
model systematically and straightforwardly up to three loops, as we demonstrate below.

The properties of the $d$=3+1 quantum XY model at zero temperature or near the critical temperature, have been addressed with spin-wave
theory, high-temperature expansions, Monte Carlo simulations, and yet other methods \citep{BEL70,BL73,Lee73,TE73,DR75,Lee75,Bet77,OB78,
RBL78,UTT79,KL79,AHN80,Bra80,AHN81,Lee83,GJ87,HM90,TFS91,WOH91,GHM93,Zha93,OHW94,TY94,GP95,BS97,SFBR01,CEHMS02}. However, regarding its
behavior at low temperatures -- between the two extremes $T$=$0$ and $T$=$T_c$ -- only little appears to be known. One result concerns the
finite-temperature susceptibility in the presence of an external field $\vec H$, that diverges as $T/\sqrt{|{\vec H}|}$ in weak
fields \citep{AHN80}.

It should be stressed that the results presented in Ref.~\citep{AHN80} -- in effective field theory language -- merely correspond to
one-loop effects. Above all, the references \citep{BEL70,BL73,Lee73,TE73,DR75,Lee75,Bet77,OB78,RBL78,UTT79,KL79,AHN80,Bra80,AHN81,Lee83,
GJ87,HM90,TFS91,WOH91,GHM93,Zha93,OHW94,TY94,GP95,BS97,SFBR01,CEHMS02} do not address the manifestation of the spin-wave interaction in
the thermodynamic behavior of the system. It is the objective of the present work to go up to three-loop order in the effective expansion
and to systematically explore the impact of the spin-wave interaction at low temperatures and weak external fields.

Our study thus extends the knowledge on the low-temperature behavior of the $d$=3+1 quantum XY model substantially. It should be pointed
out that in the context of Heisenberg ferromagnets in three spatial dimensions, more than a hundred publications on the manifestation of
the spin-wave interaction in the spontaneous magnetization have appeared (see, e.g., \citep{Dys56a,Dys56b,Zit65,Hof02,Hof11a,RPPK13,Rad15,
Hof15b} and references therein). We hence believe that we are dealing with an interesting and important question.

While the leading contribution in the free energy density of the $d$=3+1 quantum XY model is of order $T^4$, we find that subleading
corrections are of order $T^6$ (two loops) and $T^8$ (three loops). The coefficient of the $T^6$-term can easily be expressed through
microscopic quantities, since it only involves the leading-order effective constants $F$ -- where $F^2$ is the spin-wave stiffness -- and
$\Sigma_s$, i.e., the order parameter at zero temperature and infinite volume. At the three-loop level, the situation is more complicated,
as the coefficients of the various $T^8$-terms involve next-to-leading order effective constants that are {\it a priori} unknown. Still,
we can estimate their order of magnitude, which leads us to conclude that the three-loop corrections are small.

As it turns out, the spin-wave interaction is repulsive in the pressure at low temperatures and weak magnetic or staggered fields.
Remarkably, regarding the order parameter and the susceptibility, the impact of the spin-wave interaction is rather counterintuitive: if
the external field is weak, the temperature-dependent interaction contribution in the order parameter is positive, while in the
susceptibility it is negative. It should be stressed that the present effective field theory analysis is completely systematic and goes up
to three-loop order. Trying to reach the same level of accuracy with traditional microscopic methods such as spin-wave theory would be
formidable.

Our results are valid at low temperatures and in weak fields, i.e., in a regime where both parameters $T$ and $|{\vec H}|$ are small with
respect to the exchange integral $J$ that defines the natural scale of the system. Note that, on bipartite lattices, the field ${\vec H}$
can be interpreted as magnetic field coupled to the ferromagnetic quantum XY model, or equivalently, as staggered field in connection with
the antiferromagnetic XY model. We should mention, however, that our analysis goes beyond the description of quantum spin models. Our
results apply to any $d$=3+1 (pseudo-)Lorentz-invariant system that is characterized by a spontaneously broken internal symmetry O(2)
$\to$ 1.

Effective field theories based on Goldstone bosons are widely used and well established in particle physics. However, in condensed matter
physics, systematic effective Lagrangian techniques do not have the same status. But it is a fact that the low-energy behavior of many
condensed matter systems is determined by Goldstone bosons -- and the effective Lagrangian method is just designed for these systems
\citep{Leu94a,ABHV14}. I would like to point out to the condensed matter community that in the past, various systems have successfully
been analyzed within effective Lagrangian field theory. In particular, systems where the relevant Goldstone excitations are spin waves.
These comprise ferromagnetic spin chains \citep{GHKW10,Hof13a,Hof14b}, ferromagnetic films \citep{Hof12a,Hof12b,Hof14c}, and ferromagnets
in three spatial dimensions \citep{Hof11a,RPPK13,Rad15,Hof99a,RS99a,RS99b,Hof02}. They also include antiferromagnets and XY models in two
\citep{HN91,HN93,Hof10,Hof14a,Hof16a} and three \citep{HL90,Hof99b, RS00} spatial dimensions. Apart from these systems whose physics is
governed by Goldstone bosons, there are situations where additional excitations come into play: for instance high-temperature
superconductors that also involve doped holes or electrons. Systematic effective field theories have also been constructed and applied to
these remarkable systems, taking into account both square lattice \citep{KMW05,BKMPW06,BKPW06,BHKPW07,BHKMPW07,BHKPW08,JKHW09,VHJW12} and
honeycomb lattice \citep{KBWHJW12,VHJW15} geometries. Finally, we refer to Refs.~\citep{WJ94,GHJNW09,JW11,Jia11,GHJPSW11}, where the
correctness and consistency of the effective field theory method has been demonstrated in high-accuracy Monte Carlo simulations.

The organization of the paper is as follows. In Sec.~\ref{EffectiveTheory} we discuss some basic aspects of the microscopic and the
effective description of the $d$=3+1 quantum XY model. We then review the evaluation of the free energy density at low temperature up to
three-loop order in Sec.~\ref{FreeEnergyDensity}. The low-temperature expansions for the pressure, the (staggered) magnetization and
(staggered) susceptibility are obtained in Sec.~\ref{Results}. There we also discuss how the external field influences the spin-wave
interaction in these quantities. We find it instructive to compare our three-loop results for the $d$=3+1 quantum XY model with the
analogous findings for the quantum XY model in two spatial dimensions -- this is done in Sec.~\ref{Comparison}. Finally, in
Sec.~\ref{Summary} we present our conclusions. Technical details that concern the evaluation of the partition function Feynman graphs, the
numerical evaluation of a specific three-loop graph, and the estimation of subleading effective constants are presented in three separate
appendices.

\section{Effective Versus Microscopic Description}
\label{EffectiveTheory}

The microscopic Hamiltonian for the ferromagnetic $d$=3+1 quantum XY model is
\begin{equation}
\label{xyModel}
{\cal H} = - J \sum_{\langle xy \rangle} (S^1_x S^1_y + S^2_x S^2_y) - {\vec H} \cdot \sum_x {\vec S_x} \, , \qquad J > 0 \, .
\end{equation}
Here $x$ and $y$ are nearest-neighbor pairs of lattice sites separated by the distance $\hat a$, the quantity $J$ is the exchange
integral, and ${\vec H}=(0,H)$ is a weak magnetic field in the XY-plane. We assume that the crystal lattice is bipartite, such that there
is a mapping between the ferromagnetic ($J>0$) and the antiferromagnetic ($J<0$) model. Our formalism thus either describes the
ferromagnetic quantum XY model in an external magnetic field ${\vec H}$, or the antiferromagnetic quantum XY model in an external
staggered field ${\vec H_s}$ \citep{Hof14a,HOW91}. While the magnetic field is associated with the magnetization,
\begin{equation}
{\vec S} = \Big( \sum_x S^1_x, \sum_x S^2_x \Big) \, ,
\end{equation}
the staggered field couples to the staggered magnetization vector,
\begin{equation}
{\vec {\cal S}} = \Big( \sum_x (-1)^{\frac{x_1+x_2}{\hat a}} S^1_x,\sum_x (-1)^{\frac{x_1+x_2}{\hat a}} S^2_x \Big) \, .
\end{equation}
At zero temperature and infinite volume, the vacuum expectation values,
\begin{eqnarray}
\Sigma &= & \langle 0 | \sum_x S^2_x | 0 \rangle / V \, , \nonumber \\
\Sigma_s & = & \langle 0 | \sum_x (-1)^{\frac{x_1+x_2}{\hat a}} S^2_x | 0 \rangle / V \,,
\end{eqnarray}
are nonzero, signaling the spontaneous breakdown of the $O(2)$ spin rotation symmetry. In the following, we will stick to the latter
realization, bearing in mind that whenever we speak of the aniferromagnetic XY model in a staggered field, it can also be interpreted as
ferromagnetic XY model in a magnetic field.

We now leave the microscopic description and consider the $d$=3+1 quantum XY model within the effective field theory formalism. The
essential point is that the system is characterized by a spontaneously broken continuous and global spin rotation symmetry: whereas the
Hamiltonian is invariant under O(2), the ground state is not. As a consequence, a Goldstone boson excitation emerges -- a magnon or spin
wave -- that dominates the low-temperature physics of the system. Note that these low-energy degrees of freedom obey a linear, i.e.,
relativistic dispersion relation. We then define a unit vector field $U^i(x)$,
\begin{equation}
U^i(x) U^i(x) = 1 , \qquad i=1,2 ,
\end{equation}
whose first component $U^1$ corresponds to the magnon field.

The effective field theory approach is legitimate at low energies or low temperatures, and corresponds to a systematic derivative
expansion of the effective Lagrangian. Clearly, the low-energy physics of the system is dominated by terms that contain just a few time or
space derivatives. Higher-derivative terms are successively suppressed and hence are less important. Regarding the $d$=3+1 quantum XY
model, the leading piece in the effective Lagrangian is 
\begin{eqnarray}
\label{L2space}
{\cal L}^2_{eff} & = & \mbox{$ \frac{1}{2}$} F_1^2 {\partial}_0 U^i \partial_0 U^i
- \mbox{$ \frac{1}{2}$} F_2^2 \partial_r U^i \partial_r U^i + \Sigma_s H_s^i U^i \, , \qquad r = 1,2,3 \, ,
\end{eqnarray}
and involves terms with two time (${\partial}_0 {\partial}_0$) and two space (${\partial}_r {\partial}_r$) derivatives. Both contributions
are of order $p^2$, and so is the staggered field $H_s^i$. At this point we have three effective constants, $F_1, F_2$, and ${\Sigma_s}$.

Since time and space derivatives are on the same footing, the spin-wave dispersion relation can be written in a relativistic form, 
\begin{equation}
\omega = \sqrt{v^2 {\vec k}^2 + v^4 M^2} , \qquad v = \frac{F_2}{F_1} \, ,
\end{equation}
where $v$ is the spin-wave velocity, and the magnon "mass" -- or energy gap -- is identified with
\begin{equation}
\label{GBMass}
M^2 = \frac{{\Sigma}_s H_s}{F^2} \, .
\end{equation}
If one interprets the spin-wave velocity as the "velocity of light" - and furthermore sets $v \equiv 1$ -- relativistic notation can be
used, and the leading-order effective Lagrangian then takes the (pseudo-)Lorentz-invariant structure
\begin{equation}
\label{L2}
{\cal L}^2_{eff}= \mbox{$ \frac{1}{2}$} F^2 \partial_{\mu} U^i \partial^{\mu} U^i + \Sigma_s H_s^i U^i \, , \qquad F_1 = F_2 = F \, .
\end{equation}

It is important to point out that we are not dealing with an approximation here. In the effective description, as is well-known
\citep{HN93}, anisotropies due to the cubic lattice geometry only start to emerge at next-to-leading order in the derivative expansion of
the effective Lagrangian. The leading piece ${\cal L}^2_{eff}$ is strictly (pseudo-)Lorentz-invariant. On the other hand, higher-order
contributions in the effective Lagrangian do not share this accidental symmetry, and in principle all terms permitted by the discrete
symmetries of the lattice have to be considered. In the present study, however, we assume that subleading contributions in the effective
Lagrangian can also be written in a relativistic form. As we discuss in the next section, this idealization will neither affect the
general structure of the low-temperature series nor our conclusions.

The only subleading piece in the effective Lagrangian that is explicitly needed for our calculation is the next-to-leading piece
${\cal L}^4_{eff}$. Assuming (pseudo-)Lorentz-invariance, it takes the form \citep{Hof99b,HL90}
\begin{eqnarray}
\label{Leff4}
{\cal L}^4_{eff} & = & e_1 (\partial_{\mu} U^i \partial^{\mu} U^i)^2 + e_2 (\partial_{\mu} U^i \partial^{\nu} U^i)^2
+ k_1 \frac{\Sigma_s}{F^2} (H_s^i U^i) (\partial_{\mu} U^k \partial^{\mu} U^k) \nonumber \\
& & + k_2 \frac{{\Sigma}_s^2}{F^4} (H_s^i U^i)^2 + k_3 \frac{{\Sigma}_s^2}{F^4} H_s^i H_s^i \, .
\end{eqnarray}
While there are two coupling constants ($F, {\Sigma_s}$) at leading order, at next-to-leading order five additional effective constants --
$e_1, e_2, k_1, k_2$ and $k_3$ -- are required. Since the subsequent contributions ${\cal L}^6_{eff}$ (${\cal L}^8_{eff}$) only show up in
one-loop (tree) graphs, we do not need their explicit form (see next section).

The derivative structure of the terms in the effective Lagrangian is fixed by the symmetries of the underlying $d$=3+1 quantum XY
Hamiltonian. The effective constants $F, {\Sigma_s},e_1, e_2, k_1, k_2, k_3$, however, are not determined by the symmetries of the system.
One possibility to ascertain their numerical values is to match the effective calculation with the analogous microscopic calculation,
provided the latter is available -- this can be done for $F$ and $\Sigma_s$. Another possibility to fix the effective constants is by
Monte Carlo simulation or by experiment. Unfortunately, for the $d$=3+1 quantum XY model, Monte Carlo simulations, microscopic
calculations or experiments that would allow one to determine next-to-leading order (NLO) effective constants, seem to be lacking. But we
can still estimate their values.

\section{Free Energy Density up to Three-Loop Order}
\label{FreeEnergyDensity}

The perturbative evaluation of the partition function for (pseudo-)Lorentz-invariant systems that live in three spatial dimensions and
that are characterized by a spontaneously broken global symmetry O($N$) $\to$ O($N$-1), has been presented in Ref.~\citep{Hof99b}. This
effective field theory analysis was carried to three-loop order.

While our discussion is partially based on results derived in Ref.~\citep{Hof99b}, we stress that here we go much beyond this reference.
First, Ref.~\citep{Hof99b} focused on the Heisenberg antiferromagnet, and not on the quantum XY model where new subtleties in the
renormalization process and the structure of the low-temperature expansion occur. Then, we numerically evaluate a rather complicated
three-loop diagram. Moreover, the discussion in Ref.~\citep{Hof99b} concerned the limit of a zero external field, while here we explore
the thermodynamics in presence of a weak external field. Finally, the susceptibility has not been considered in Ref.~\citep{Hof99b}.

In the present section, and in appendix \ref{appendixA}, we provide a concise review of the basic formulas obtained in Ref.~\citep{Hof99b}.
The evaluation of the partition function within effective field theory has been outlined in detail in Ref.~\citep{Hof10} (section 2) and
Ref.~\citep{Hof11a} (appendix A), much beyond the sketch we give below. The interested reader is referred to these references and to
Refs.~\citep{Brau10,Bur07,Leu95} that provide pedagogic introductions to the effective Lagrangian method.

\begin{figure}[t]
\begin{center}
\includegraphics[width=16cm]{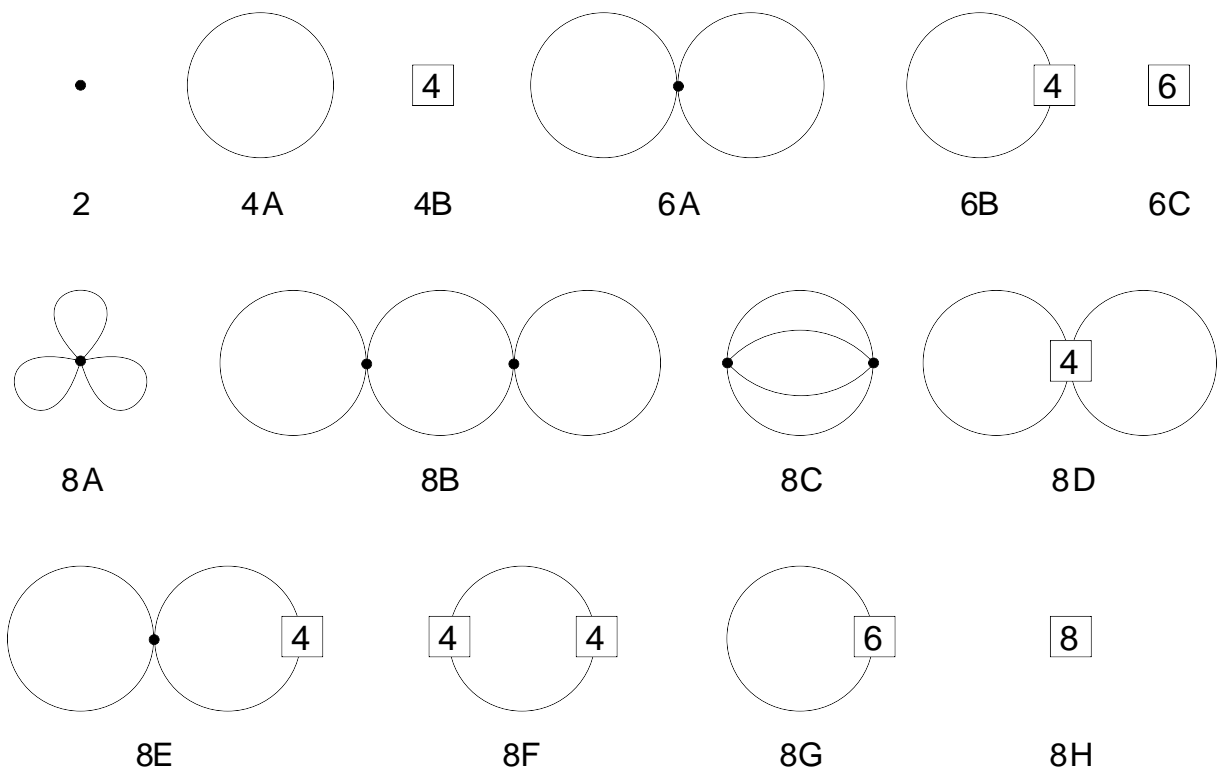}
\end{center}
\caption{$D$=3+1 quantum XY model: Feynman diagrams occurring in the low-temperature expansion of the partition function up to three-loop
order $T^8$. Vertices involving the leading piece ${\cal L}^2_{eff}$ of the effective Lagrangian correspond to a filled circle, while
vertices associated with ${\cal L}^4_{eff},{\cal L}^6_{eff}, {\cal L}^8_{eff}$ are denoted by the numbers $4,6,8$, respectively. Each loop
is suppressed by $p^2 \propto T^2$.}
\label{figure1}
\end{figure}

The perturbative expansion of the partition function is based on the observation that each Goldstone-boson loop is suppressed by some
power $n$ of temperature. The number $n$ depends both on the spatial dimension $d_s$ of the system and on the nature of the dispersion
relation of its Goldstone bosons. Provided that the dispersion relation is linear, the suppression is $p^{d_s-1} \propto T^{d_s-1}$
\citep{HL90}. In the present case of the $d$=3+1 quantum XY model, each loop is suppressed by two powers of momentum or temperature. This
provides the basis to understand the organization of the Feynman diagrams of Fig.~\ref{figure1} that are relevant up to three-loop order.

The leading temperature-dependent term comes from diagram 4A, yielding a contribution of order $p^4 \propto T^4$ in the free energy
density: this is the free Bose gas term. We then have a two-loop contribution (graph 6A) of order $p^6 \propto T^6$, and various
three-loop contributions (graphs 8A-C) of order $p^8 \propto T^8$. Note that tree-level diagrams just correspond to $T$-independent
contributions that can be absorbed into the vacuum energy density. The only $T$-dependent contribution that requires a piece of the
effective Lagrangian beyond ${\cal L}^4_{eff}$, is the one-loop diagram 8G that involves ${\cal L}^6_{eff}$. However, this graph merely
contributes to the renormalization of the Goldstone boson mass (see appendix \ref{appendixA}).

\begin{figure}[t]
\includegraphics[width=15cm]{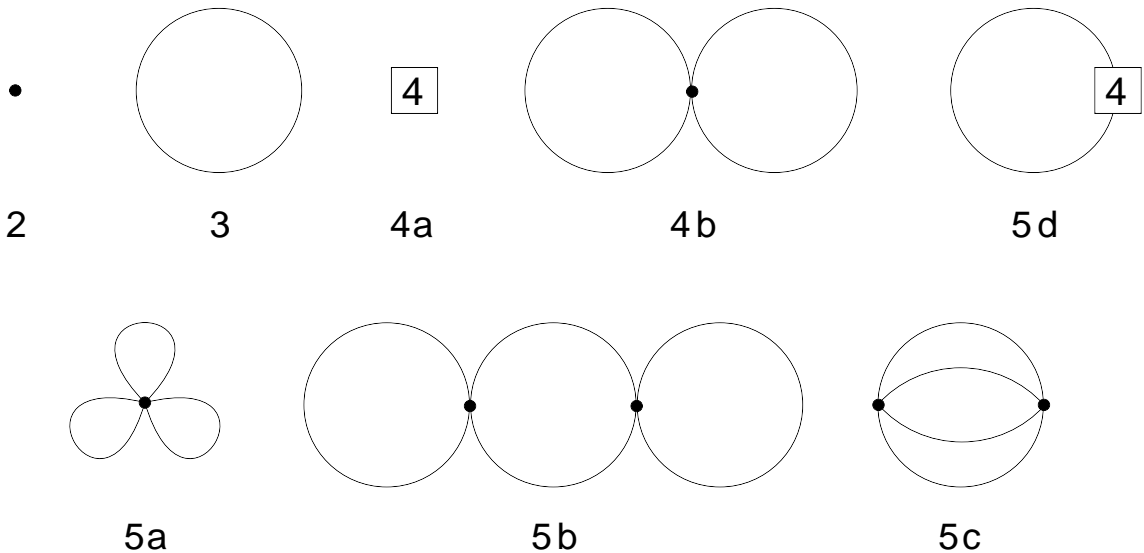}
\caption{$D$=2+1 quantum XY model: Feynman diagrams occurring in the low-temperature expansion of the partition function up to three-loop
order $T^5$. Vertices involving the leading piece ${\cal L}^2_{eff}$ of the effective Lagrangian correspond to a filled circle, while
vertices associated with ${\cal L}^4_{eff}$ are denoted by the number $4$. Each loop is suppressed by $p \propto T$.}
\label{figure2}
\end{figure}

In three spatial dimensions, the loop-suppression by $T^2$ leads to the general pattern $T^4, T^6, T^8$ in the free energy density.
Regarding the quantum XY model in two spatial dimensions, the situation is different since each loop is only suppressed by one power of
temperature. This leads to the Feynman diagrams depicted in Fig.~\ref{figure2}. Here the leading $T$-dependent term comes from diagram 3:
the free Bose gas contribution of order $p^3 \propto T^3$. The two-loop (three-loop) corrections then yield terms of order $ T^4$ ($T^5$)
in the free energy density. Further differences and analogies between the quantum XY model in three and two spatial dimensions are
presented in Sec.~\ref{Comparison}.

As described in appendix \ref{appendixA}, up to order $p^8 \propto T^8$ in the free energy density, the final result for general $N \ge 2$
amounts to
\begin{equation}
\label{freeED}
z = z_0 - \mbox{$ \frac{1}{2}$} (N\!-\!1) g_0 - 4 \pi a (g_1)^2 - \pi g \, \Big[ b - \frac{j}{{\pi}^3 F^4} \Big] + {\cal O}(p^{10}) \, .
\end{equation}
This representation applies to any $d$=3+1 (pseudo-)-Lorentz-invariant system exhibiting the spontaneous symmetry breaking pattern O($N$)
$\to$ O($N$-1). The various quantities are defined in appendix \ref{appendixA}. These include the $T$=0 free energy density $z_0$
(\ref{vacuumEnergyDensity}), the parameters $a$ and $b$ that contain NLO effective constants (\ref{ConstAB}), the kinematical functions
$g_r$ (\ref{FreeFunctions}) and $g$ (\ref{gCombSimpl}) that nontrivially depend on temperature and on the renormalized magnon mass
$M_{\pi}$. The latter can be expressed in terms of the staggered field $H_s$ (\ref{renMass}). Finally, $j$ is a dimensionless function
defined in (\ref{functIJ}) -- much like the kinematical functions $g_r$ and $g$, it depends on the dimensionless ratio $\tau$,
\begin{equation}
\tau = \frac{T}{M_{\pi}} .
\end{equation}
The numerical evaluation of the three-loop integral $j$ is outlined in appendix \ref{appendixB}.

The structure of the low-temperature expansion becomes more transparent if the following dimensionless functions $h_0, h_1$ and $h$ are
introduced:
\begin{equation}
\label{ThermalDimensionless}
g_0(\sigma) = T^4 h_0(\sigma) , \qquad g_1(\sigma) = T^2 h_1(\sigma) , \qquad g(\sigma) = T^8 h(\sigma) \, ,
\end{equation}
where the dimensionless parameter $\sigma$ is
\begin{equation}
\sigma = \frac{M_{\pi}}{2 \pi T} = \frac{1}{2 \pi \tau} \, .
\end{equation}

\begin{figure}[t]
\begin{center}
\includegraphics[width=14cm]{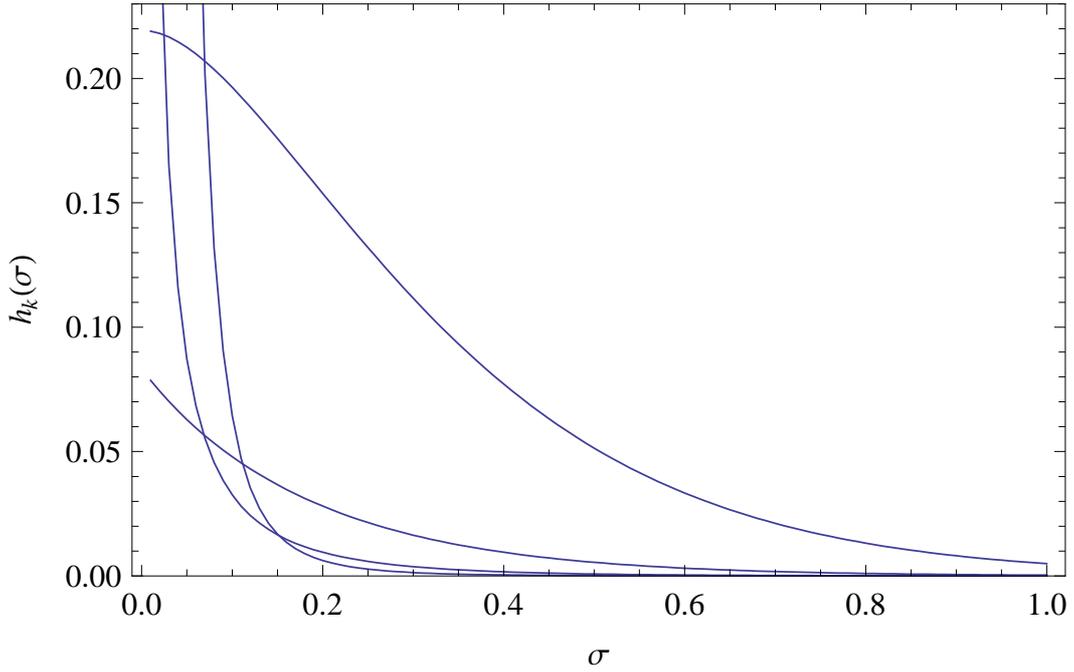}
\end{center}
\caption{The kinematical functions $h_0(\sigma), h_1(\sigma), h_2(\sigma), h_3(\sigma)$ from top to bottom in the figure (vertical cut at
$\sigma = 0.2$), as a function of the parameter $\sigma = M_{\pi}/2\pi T$.}
\label{figure4}
\end{figure}

The kinematical functions $h_0, h_1, h_2, h_3$ are depicted in Fig.~\ref{figure4}. The latter two are relevant in the magnetization and
susceptibility -- they scale like 
\begin{equation}
g_2(\sigma) = h_2(\sigma) , \qquad g_3(\sigma) = \frac{h_3(\sigma)}{T^2} \, .
\end{equation}

Using these representations, the low-temperature expansion for the free energy density of the $d$=3+1 quantum XY model takes the form
\begin{eqnarray}
\label{fedTauXY}
z & = & z_0 
- \mbox{$ \frac{1}{2}$} h_0(\sigma) T^4
- \frac{1}{8 F^2 t^2} {h_1(\sigma)}^2 T^6
- \frac{3(e_1 + e_2) + \mbox{$ \frac{1}{2}$} {\overline k} - \mbox{$ \frac{3}{256 \pi^2}$}}{F^4 t^4} \, {h_1(\sigma)}^2
\, T^8 \nonumber \\
& & - \frac{2(e_1 + e_2)}{F^4} \, h(\sigma) \, T^8
+ \frac{1}{\pi^2 F^4} \, j(\sigma) h(\sigma) \, T^8 + {\cal O}(p^{10})\, , \qquad (N=2)
\end{eqnarray}
where $t$ is the dimensionless ratio 
\begin{equation}
t = \frac{T}{M} = \frac{T F}{\sqrt{\Sigma_s H_s}} \, .
\end{equation}
Up to two-loop order $T^6$, the coefficients merely involve the effective constants $F$ and $\Sigma_s$ from ${\cal L}^2_{eff}$. NLO
effective constants start showing up at three-loop order $T^8$: for the definition of $e_1, e_2$ and ${\overline k}$ see Eq.~(\ref{Leff4})
and Eq.~(\ref{renormk1k2}).\footnote{Note that we use two-loop order (three-loop order) synonymous for order $T^6$ ($T^8$), since in our
convention we only count loop-graphs that exclusively contain insertions from ${\cal L}^2_{eff}$. It should be kept in mind that at order
$T^6$ we also have a one-loop graph (6B), and that at order $T^8$ we have also have two-loop (8D,E) and one-loop (8F,G) graphs (see
Fig.~\ref{figure1}).} The low-temperature expansion is characterized by even powers of the temperature. The free Bose gas term (order
$T^4$) receives corrections of ascending powers $T^2$.

If one aims at three-loop accuracy, it is important to distinguish between $t$ and $\tau$: whereas the former involves the leading-order
mass $M$, the latter involves the renormalized mass $M_{\pi}$ (see Eq.~(\ref{renMass})). The difference between $1/t^2$ and $1/\tau^2$ is
of order $M^4$. If one is only interested in two-loop accuracy, it is hence legitimate to replace $1/t^2$ by $1/\tau^2$ in the
contribution of order $T^6$.

Here is the appropriate place to elaborate on (pseudo-)Lorentz-invariance where our approach is based upon. Abandoning
(pseudo-)Lorentz-invariance can be done in two steps: (i) write down all terms that are still consistent with space-rotation invariance,
or (ii) even abandon the idealization of space isotropy, by taking into account all terms that are consistent with the discrete symmetries
of the lattice. How would the low-temperature series be affected? The point is that the temperature powers would not change, but the
coefficients would be different: more subleading effective constants would show up. Also, it would no longer be possible to define a
Goldstone boson mass $M_{\pi}$ via the relativistic dispersion relation
\begin{equation}
\omega = \sqrt{v^2 {\vec k}^2 + v^4 M_{\pi}^2} \, .
\end{equation}
It is not our intention, however, to explicitly incorporate all these subleading effective constants. Remember that Lorentz-symmetry
breaking terms only start emerging in ${\cal L}^4_{eff}$. According to Fig.~\ref{figure1}, the temperature-dependent interaction is thus
only affected through the two-loop diagrams 8D and 8E, i.e., at next-to-next-to-leading order $T^8$ in the free energy density. For
practical purposes it is legitimate to work within a (pseudo-)Lorentz-invariant formalism. After all, we are dealing with small effects,
as we illustrate below.

\section{Low-Temperature Series}
\label{Results}

The low-temperature representation for the free energy density, Eq.~(\ref{fedTauXY}), provides the basis for our subsequent discussion --
any other thermodynamic observable can be derived from there. But let us first clarify in which parameter range -- defined by temperature
and external field -- our series are valid. The effective expansion is restricted to low energies. A natural energy scale is the exchange
constant $J$ inherent in the underlying microscopic model, such that our series are valid as long as both $T$ and $H_s$ are small compared
to $J$. Equivalently, we may consider the critical temperature $T_c$ where the order parameter drops to zero and the phase transition
takes place -- at or near this point, the spin-wave picture is no longer adequate. According to Ref.~\citep{BEL70}, for the simple-cubic
quantum XY model one has $T_c \approx 2.02 J$, and one may define low temperature and weak field as
\begin{equation}
T, H_s, M_{\pi} \ \lesssim \ 0.2 \ T_c \ \approx \ 0.4 \, J \, .
\end{equation}
Furthermore, as we outline in appendix \ref{appendixC}, the connection between the microscopic scale $J$ and the effective constant $F$ is
approximately $F \approx 0.4 J$ for the quantum XY model on the simple cubic lattice. Accordingly, the parameter ranges translate into
\begin{equation}
T, H_s, M_{\pi} \ \lesssim \ F \, .
\end{equation}

We now consider the low-temperature series for the pressure, order parameter, and susceptibility. One interesting topic is to explore the
sign and strength of the spin-wave interaction in these quantities as a function of temperature and external field.

\subsection{Pressure}
\label{ResultsPressure}

The pressure can be extracted from the free energy density through
\begin{equation}
P = z_0 - z \, .
\end{equation}
For the $d$=3+1 quantum XY model we obtain
\begin{eqnarray}
\label{fedTau}
P(T,H_s) & = & \mbox{$ \frac{1}{2}$} h_0(\sigma) T^4
+ \frac{1}{8 F^2 t^2} {h_1(\sigma)}^2 T^6
+ \frac{3(e_1 + e_2) + \mbox{$ \frac{1}{2}$} {\overline k} - \mbox{$ \frac{3}{256 \pi^2}$ }}{F^4 \tau^4} \, {h_1(\sigma)}^2 \, T^8
\nonumber \\
& & + \frac{2(e_1 + e_2)}{F^4} \, h(\sigma) \, T^8
- \frac{1}{\pi^2 F^4} \, j(\sigma) h(\sigma) \, T^8 + {\cal O}(T^{10}) \, .
\end{eqnarray}
The spin-wave interaction shows up at order $T^6$, subsequent corrections are of order $T^8$. If the external field is switched off, the
pressure reduces to
\begin{equation}
P(T,0) = \frac{\pi^2}{90} \, T^4 + \frac{2 \pi^4 (e_1 + e_2)}{675 F^4} \, T^8 \, .
\end{equation}
There is no $T^6$-contribution in this case.

\begin{figure}[t]
\includegraphics[width=13.5cm]{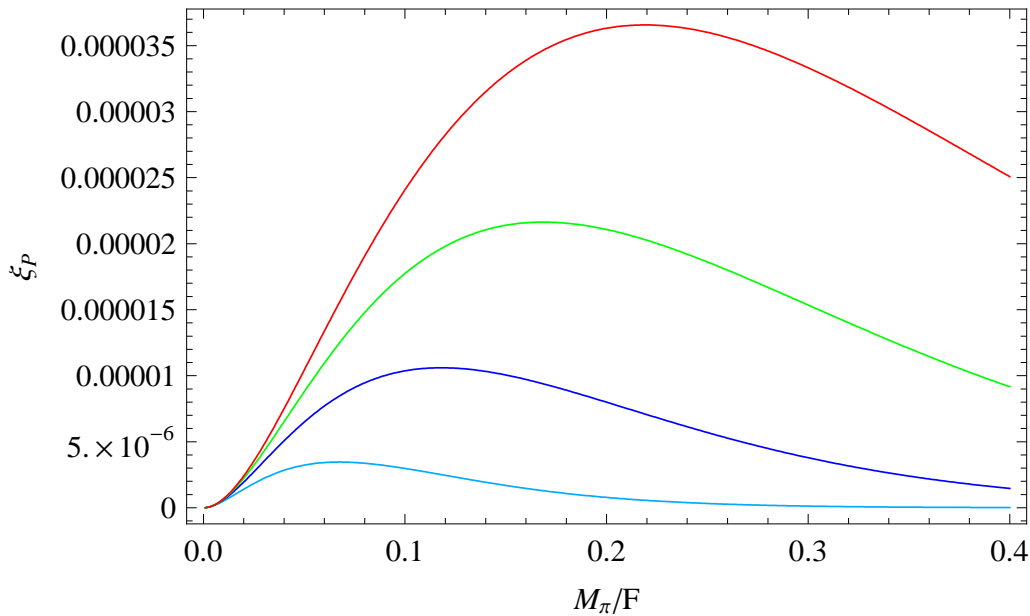}
\caption{[Color online] Manifestation of the leading interaction contribution in the pressure of the $d$=3+1 quantum XY model, measured by
$\xi_P$. The curves refer to the temperatures $T/F = \{ 0.04, 0.07, 0.10, 0.13 \}$ from bottom to top in the figure.}
\label{figure5}
\end{figure}

In Fig.~\ref{figure5} we compare the strength of the dominant interaction correction ($T^6$) with the free Bose gas term ($T^4$), using
the ratio
\begin{equation}
\label{intRatioP}
\xi_P(T,H_s) = \frac{P^6_{int}(T,H_s)}{P_{Bose}(T,H_s)}
\end{equation}
for the temperatures $T/F = \{ 0.04, 0.07, 0.10, 0.13 \}$ in presence of a weak external field, parametrized by $M_{\pi}/F$.
The sign of $\xi_P$ is positive, meaning that the spin-wave interaction in the pressure is repulsive, irrespective of the temperature and
strength of the external field. The interaction is maximal in an intermediate domain of staggered field strength. The corresponding maxima,
however, are tiny -- the spin-wave interaction in the $d$=3+1 quantum XY model is very weak also in nonzero external field.

At order $T^8$ there are three terms that contribute to the interaction. The coefficients of the first two terms involve NLO effective
constants that are a priori unknown. They could be determined by matching our formulas with corresponding microscopic formulas, by Monte
Carlo simulations of the $d$=3+1 quantum XY model at low temperatures, or by comparing our predictions with experiments. Unfortunately,
none of these options seem to be available.

We can, however, estimate the order of magnitude of these NLO effective constants. The explicit steps can be found in appendix
\ref{appendixC}. The outcome is 
\begin{equation}
|e_1| \approx |e_2| \approx |{\overline k}| \approx 0.001 \, .
\end{equation}
While the numerical values of these NLO effective constants are small, unfortunately, their sign still is inconclusive. It should be noted
that the effective constant $\overline k$ in Eq.~(\ref{fedTau}), unlike $e_1$ and $e_2$, requires logarithmic renormalization (see
appendix \ref{appendixA}).

\begin{figure}[t]
\includegraphics[width=13.5cm]{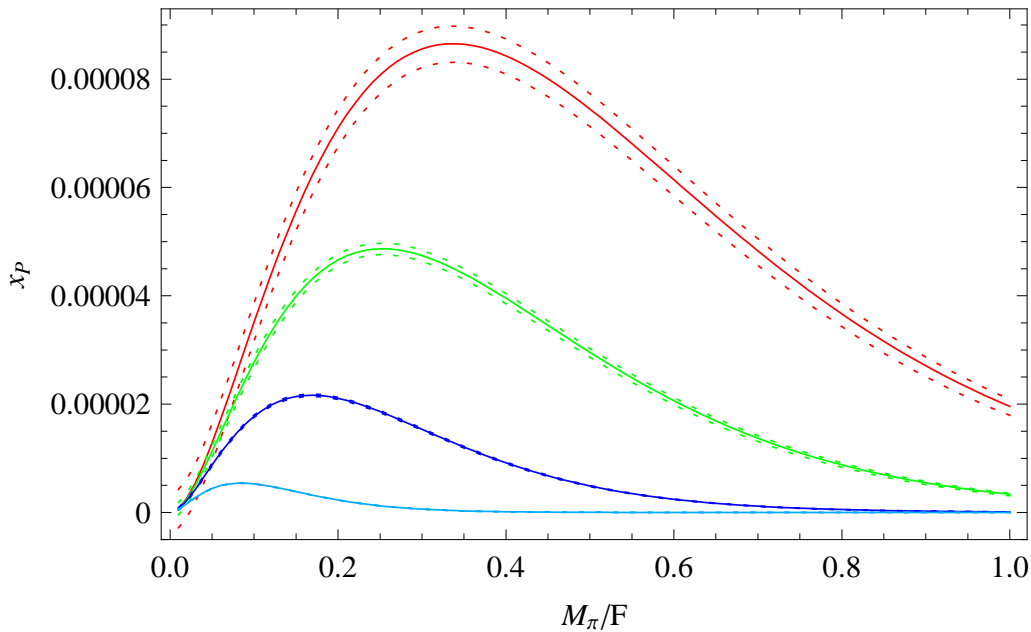}
\caption{[Color online] Spin-wave interaction in the pressure of the $d$=3+1 quantum XY model: the three-loop correction is very small
compared to the two-loop contribution, as measured by $x_P$. The curves refer to the temperatures $T/F = \{ 0.05, 0.10, 0.15, 0.20 \}$
from bottom to top in the figure.}
\label{figure6}
\end{figure}

In Fig.~\ref{figure6} we display the ratio
\begin{equation}
x_P(T,H_s) = \frac{P^8_{int}(T,H_s) + P^6_{int}(T,H_s)}{P_{Bose}(T,H_s)}
\end{equation}
that measures the strength of the $T^8$- and $T^6$-interaction correction with respect to the Bose term. We depict two extreme scenarios:
on the one hand, $e_1,e_2$ and ${\overline k}$ all taking positive values of the order $0.001$, and, one the other hand $e_1 \approx e_2
\approx {\overline k} \approx -0.001$. Compared to the previous figure, we have chosen higher temperatures, $T/F = \{ 0.05, 0.10, 0.15,
0.20 \}$, and larger values for the staggered field. One notices that the $T^8$-correction may be positive or negative, signaling that the
repulsive $T^6$-contribution may be enhanced or weakened. In very weak staggered fields, the interaction among spin waves may even become
attractive -- but note that the strength of the interaction in this region is tiny. In conclusion, it is justified to only consider the
leading correction of order $T^6$. In the ensuing plots, we will indeed restrict ourselves to two-loop accuracy.

Regarding the interpretation of the figures that show the effect of the spin-wave interaction, it is important to point out a subtlety
related to $M$ and $M_{\pi}$. The crucial point is that the spin-wave interaction also manifests itself in temperature-independent
quantities, in particular, in Eq.~(\ref{renMass}) that connects $M_{\pi}$ and $M$: the second (third) term on the RHS is the two-loop
(three-loop) correction that the leading term $M^2$ receives -- both terms originate from the spin-wave interaction at $T$=0. What we
have depicted in the figures is the {\it finite-temperature} interaction contribution that depends on $M_{\pi}$ which incorporates the
interaction at $T$=0. The interpretation of the figures presented in this subsection -- and those presented below -- hence is as follows:
We start at zero temperature and switch on the external field. Afterwards we go from $T$=0 to finite temperature -- while keeping $H_s$
fixed -- and study how this affects the interaction at {\it finite temperature}. In the pressure, the interaction is repulsive.

\subsection{Order Parameter}
\label{ResultsOrderParameter}

The staggered magnetization at finite temperature is defined by
\begin{equation}
\Sigma_s(T,H_s) = - \frac{\partial z}{\partial H_s} \, .
\end{equation}
This is the order parameter, signaling that the internal symmetry O(2) is spontaneously broken.

With the representation for the free energy density, Eq.~(\ref{fedTauXY}), the low-temperature series for the staggered magnetization
amounts to 
\begin{eqnarray}
\label{OPTauN2}
\Sigma_s(T,H_s) & = & \Sigma_s(0,H_s) - \frac{\Sigma_s {\hat b}}{2 F^2} h_1(\sigma) \, T^2
+ \frac{\Sigma_s}{8 F^4} \Big\{ {h_1(\sigma)}^2 - \frac{2 {\hat b}}{t^2} h_1(\sigma) h_2(\sigma) \Big\} \, T^4 \nonumber \\
& + & \frac{2 \Sigma_s}{t^2 F^6} \Big\{ 3(e_1 + e_2) + \mbox{$ \frac{1}{2}$} {\overline k} - \mbox{$ \frac{3}{256 \pi^2}$} \Big\}
\Big\{ {h_1(\sigma)}^2 - \frac{\hat b}{t^2} \, h_1(\sigma) h_2(\sigma) \Big\} \, T^6 \nonumber \\
& - & \frac{3 \Sigma_s {\hat b}}{F^6} \Big\{ 2(e_1 + e_2) - \frac{1}{\pi^2}\, j(\sigma) \Big\}
\Big\{ h_0(\sigma) h_1(\sigma) + \frac{{h_1(\sigma)}^2 + h_0(\sigma) h_2(\sigma) }{\tau^2} \Big\} \, T^6 \nonumber \\
& - & \frac{3 \Sigma_s {\hat b}}{8 \pi^4 F^6 \sigma} \frac{\partial j(\sigma)}{\partial \sigma} \, \Big\{ {h_0(\sigma)}^2
+ \frac{ h_0(\sigma) h_1(\sigma) }{\tau^2} \Big\} \, T^6 \nonumber \\
& - & \frac{\Sigma_s}{64 \pi^2 F^6 t^2} \, {h_1(\sigma)}^2 \, T^6 + {\cal O}(T^8) \, .
\end{eqnarray}
The quantity $\hat b$ is
\begin{equation}
\label{defb}
{\hat b}(H_s) = \frac{\partial M^2_{\pi}}{\partial M^2} = 1 - \frac{1}{32 \pi^2} \frac{\Sigma_s H_s}{F^4}
+ 2 {\overline k} \frac{\Sigma_s H_s}{F^4} + {\cal O}(H_s^2) \, .
\end{equation}
The spin-wave interaction manifests itself at order $T^4$ and $T^6$, both in presence or absence of the staggered field.

At zero temperature, the staggered magnetization becomes
\begin{equation}
\label{SigmaN}
\Sigma_s(0,H_s) = \Sigma_s \Big\{ 1 + {\overline k} \frac{\Sigma_s H_s}{F^4} - \frac{1}{64 \pi^2} \frac{\Sigma_s H_s}{F^4} \Big\}
+ {\cal O}(H_s^2) \, ,
\end{equation}
where
\begin{equation}
\Sigma_s = \Sigma_s(0,0) \, .
\end{equation}

\begin{figure}[t]
\includegraphics[width=13.5cm]{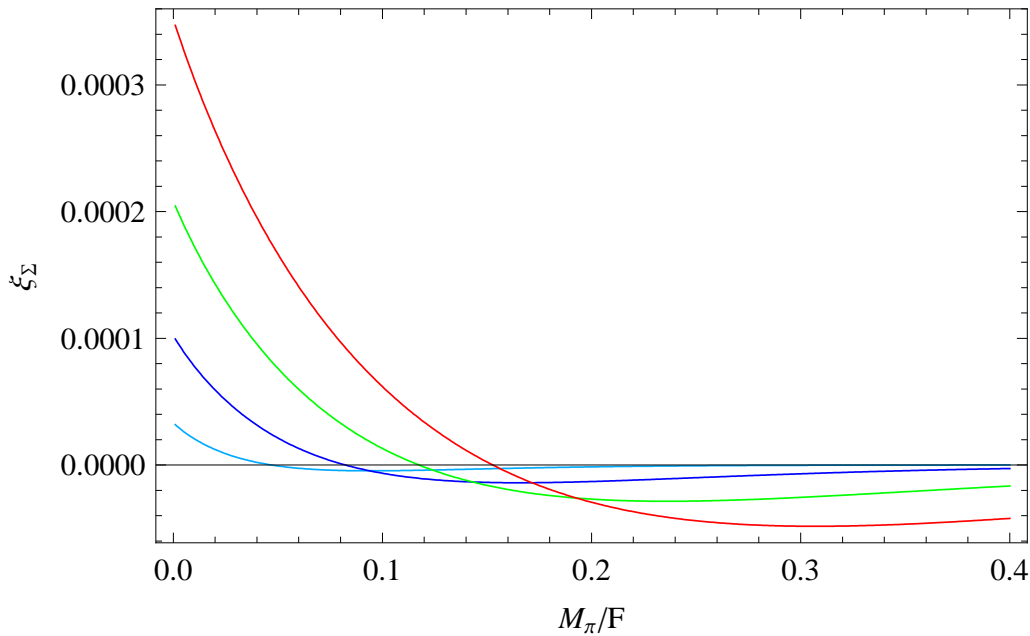}
\caption{[Color online] Manifestation of the leading interaction contribution in the staggered magnetization of the $d$=3+1 quantum XY
model, measured by $\xi_{\Sigma}$. The curves refer to the temperatures $T/F = \{ 0.04, 0.07, 0.10, 0.13 \}$ from bottom to top in the
figure (vertical cut at $M_{\pi}/F = 0$).}
\label{figure7}
\end{figure}

In Fig.~\ref{figure7} we plot the ratio
\begin{equation}
\xi_{\Sigma}(T,H_s) = \frac{{\Sigma}^4_{int}(T,H_s)}{|{\Sigma}_{Bose}(T,H_s)|} \, ,
\end{equation}
that measures strength and sign of the spin-wave interaction in the leading interaction term ($T^4$) with respect to the free Bose gas
term ($T^2$). While the quantity $\xi_{\Sigma}$ is mainly negative in parameter space, interestingly, for very weak external fields (small
$M_{\pi}$) it becomes positive.

It should be noted that these two-loop effects are quite small and that the properties of the order parameter are dominated by the
(one-loop) free Bose gas contribution. As expected, this term ($\propto T^2$) is negative: the order parameter gradually decreases as the
temperature rises. It is quite remarkable -- or counterintuitive -- that the spin-wave interaction not necessarily presents this behavior.
If the temperature is low and the field $H_s$ is weak, the behavior is just the opposite: the interaction induces an increase of the order
parameter, i.e., it tends to reinforce the (anti-)alignment of the spins and enhances the (staggered) magnetization. Keep in mind that we
first switch on the field $H_s$ at zero temperature, and then go to finite $T$ while keeping $H_s$ fixed.

\subsection{Staggered Susceptibility}
\label{ResultsSusceptibility}

The staggered susceptibility corresponds to the derivative of the order parameter, Eq.~(\ref{OPTauN2}), with respect to the staggered
field,
\begin{equation}
\label{definitionSusceptibility}
\chi(T,H_s) = \frac{\partial \Sigma_s(T,H_s)}{\partial H_s} = \frac{\Sigma_s}{F^2} \, \frac{\partial \Sigma_s(T,M)}{\partial M^2} \, .
\end{equation}
The low-temperature series exhibits the general structure 
\begin{equation}
\label{seriesSuscept}
\chi(T,H_s) = \chi(0,H_s) + \chi_1(\tau) + \chi_2(\tau) T^2 + \chi_3(\tau) T^4 + {\cal O}(T^6) \, .
\end{equation}
Since the expressions for the coefficients are rather lengthy, we do not display the full three-loop result that can trivially be obtained
from Eq.~(\ref{definitionSusceptibility}). Here we provide explicit expressions up to two-loop order and furthermore work within the
approximation ${\hat b} \approx 1$ (see Eq.~(\ref{defb})), which is sufficient for practical purposes. We then obtain 
\begin{eqnarray}
\chi_1(\tau) & = & \frac{\Sigma_s^2}{2 F^4} h_2 \, , \nonumber \\
\chi_2(\tau) & = & - \frac{\Sigma_s^2}{4 F^6} \Big\{ 2 h_1 h_2 - \frac{1}{t^2} \Big( h_2^2 + h_1 h_3 \Big) \Big\} \, .
\end{eqnarray}

\begin{figure}[t]
\includegraphics[width=13.5cm]{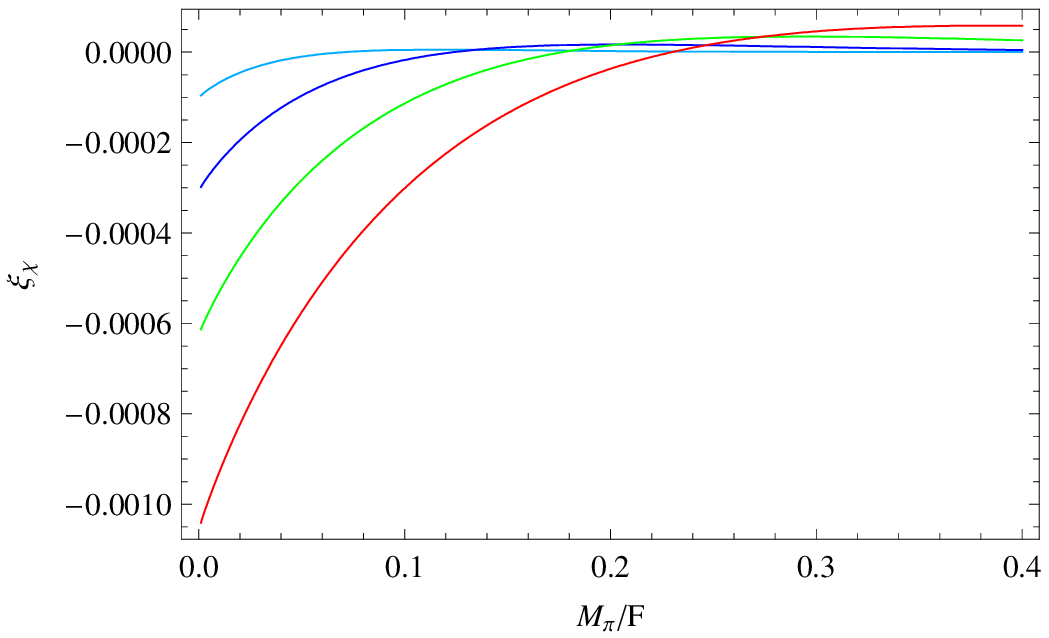}
\caption{[Color online] Manifestation of the leading interaction contribution in the staggered susceptibility of the $d$=3+1 quantum XY
model, measured by $\xi_{\chi}$. The curves refer to the temperatures $T/F = \{ 0.04, 0.07, 0.10, 0.13 \}$ from top to bottom in the
figure (vertical cut at $M_{\pi}/F = 0$).}
\label{figure8}
\end{figure}

In Fig.~\ref{figure8} we plot the ratio
\begin{equation}
\xi_{\chi}(T,H_s) = \frac{{\chi}^2_{int}(T,H_s)}{{\chi}_{Bose}(T,H_s)} \, ,
\end{equation}
that measures strength and sign of the spin-wave interaction in the leading interaction term ($T^2$) in the staggered susceptibility
relative to the free Bose gas term ($T^0$).

Remarkably, $\xi_{\chi}$ takes negative values if the staggered field is weak. Since the staggered field tends to reinforce the staggered
spin pattern and to enhance the order parameter, one would expect $\chi(T,H_s)$ to take positive values. While this is indeed the case for
the free Bose gas contribution, the effects concerning the spin-wave interaction are peculiar: negative values of $\xi_{\chi}$ imply that
the staggered field actually perturbs the antialignment of the spins, such that the order parameter decreases. It should be kept in mind,
however, that we are dealing with rather weak effects -- the free Bose gas term dominates and the overall staggered susceptibility is
positive.

In the limit $H_s \to 0$, the leading temperature-dependent term in the staggered susceptibility diverges according to
\begin{equation}
\label{divergenceChi}
\lim_{H_s \to 0} \chi_1 \propto \frac{T}{\sqrt{H_s}} \, .
\end{equation}
This corresponds to one of the few known results for the $d$=3+1 quantum XY model at low temperatures where the spin-wave picture applies
\citep{AHN80}. Note that the divergence originates from the free Bose gas term -- in this sense it is a trivial one-loop result.

\section{Quantum XY Model: $d$=3+1 Versus $d$=2+1}
\label{Comparison}

The organization of the loop expansion, as we have discussed in Sec.~\ref{FreeEnergyDensity}, depends on the spatial dimension. The
respective Feynman graphs are presented in Fig.~\ref{figure1} (three space dimensions) and Fig.~\ref{figure2} (two space dimensions). In
the case of the $d$=2+1 quantum XY model, the interaction diagrams only involve the leading piece ${\cal L}^2_{eff}$ of the effective
Lagrangian. Regarding the $d$=3+1 quantum XY model, however, we have two-loop diagrams (diagrams 8D,E in Fig.~\ref{figure1}) that contain
vertices from ${\cal L}^4_{eff}$. Accordingly, the coefficients of the interaction terms of order $T^8$ in the free energy density involve
NLO effective constants. Note that, in $d$=2+1, these two-loop diagrams are of order $T^6$, i.e., beyond three-loop level. One thus
realizes that the spontaneously broken O(2) symmetry is more restrictive in two spatial dimensions, in the sense that less effective
constants are required, i.e., less information on the specific properties of the underlying microscopic XY model is needed.

Since loops are suppressed by only one power of temperature in two spatial dimensions, the corrections to the Bose term (order $T^3$ in
the free energy density) proceed in steps of $T$. As an example we quote the representation for the pressure of the $d$=2+1 quantum XY
model (for details see Ref.~\citep{Hof14a}),
\begin{eqnarray}
P(T,H_s) & = & \mbox{$\frac{1}{2}$} h_0(\sigma) T^3
+ \frac{1}{8 F^2 {\tau}^2} {h_1(\sigma)}^2 T^4
+ \frac{1}{128 \pi F^4 {\tau}^3} {h_1(\sigma)}^2 T^5
- \frac{1}{48 F^4 {\tau}^2} {h_1(\sigma)}^3 T^5 \nonumber \\
& & + \frac{1}{16 F^4 {\tau}^4} {h_1(\sigma)}^2 h_2(\sigma) T^5
- \frac{1}{F^4} q(\sigma) T^5 + {\cal O}(T^6) \, .
\end{eqnarray}
It should be noted that the kinematical functions $h_r$, defined in Eq.~(\ref{ThermalDimensionless}) and Eq.~(\ref{FreeFunctions}),
depend on the spatial dimension. While Fig.~\ref{figure4} refers to $d$=3+1, analogous plots for the kinematical functions in $d$=2+1 are
depicted in Fig.~1 of Ref.~\citep{Hof16a}. A graph for the three-loop function $q$ is given in Fig.~3 of Ref.~\citep{Hof14a}.

The leading interaction contribution in the pressure -- both in three and two spatial dimensions -- is the two-loop term proportional to
$h^2_1$, its positive sign indicating that the interaction among the spin waves is repulsive. Regarding the order parameter, the relevant
combination at two-loop order is
\begin{equation}
\label{signMag}
h_1(\sigma)^2 - 8 {\pi}^2 {\sigma}^2 h_1(\sigma) h_2(\sigma) \, ,
\end{equation}
implying that the sign of the interaction contribution in the staggered magnetization may be negative, positive, or zero. The sign change
for $d$=2+1 occurs at $\sigma \approx 0.11$, while for $d$=3+1 it happens at $\sigma \approx 0.19$. In either case the spin-wave
interaction contribution is positive in weak staggered fields which seems counterintuitive. Clearly, this is a subtle two-loop effect, and
the dominant one-loop term always wins: the order parameter decreases if we switch on the temperature while keeping $H_s$ fixed -- as one
would expect. Similar effects can be observed in the staggered susceptibility: in weak external fields -- both in $d$=3+1 and $d$=2+1 -- it
takes negative values, indicating that the interaction perturbs the antialignment of the spins. Still, the overall staggered
susceptibility is always positive since the free Bose gas contribution dominates.

Regarding the temperature scale where the spin-wave picture ceases to be valid, we note the following. The $d$=2+1 quantum XY model is
characterized by the Kosterlitz-Thouless transition temperature that marks this point. Specifically, on the square lattice, according to
Ref.~\citep{HK98}, we have $T_{KT} \approx 0.343 J$. Now for the $d$=3+1 quantum XY model, the relevant scale is given by the transition
temperature where the staggered magnetization drops to zero. This happens at $T_c= 2.02 J$ for the simple cubic lattice \citep{BEL70}. In
units of the respective exchange integrals $J$, the temperature scales thus differ by about a factor of six. Regarding the leading-order
effective constant $F$, on the square lattice, Monte Carlo simulations yield $F^2 = 0.26974(5) J$ \citep{GHJPSW11}, which is slightly less
than the Kosterlitz-Thouless transition temperature. For the simple-cubic lattice, with the estimate (\ref{estimateFJ}), we obtain
$F \approx 0.4 J$.

\begin{figure}
\includegraphics[width=13.5cm]{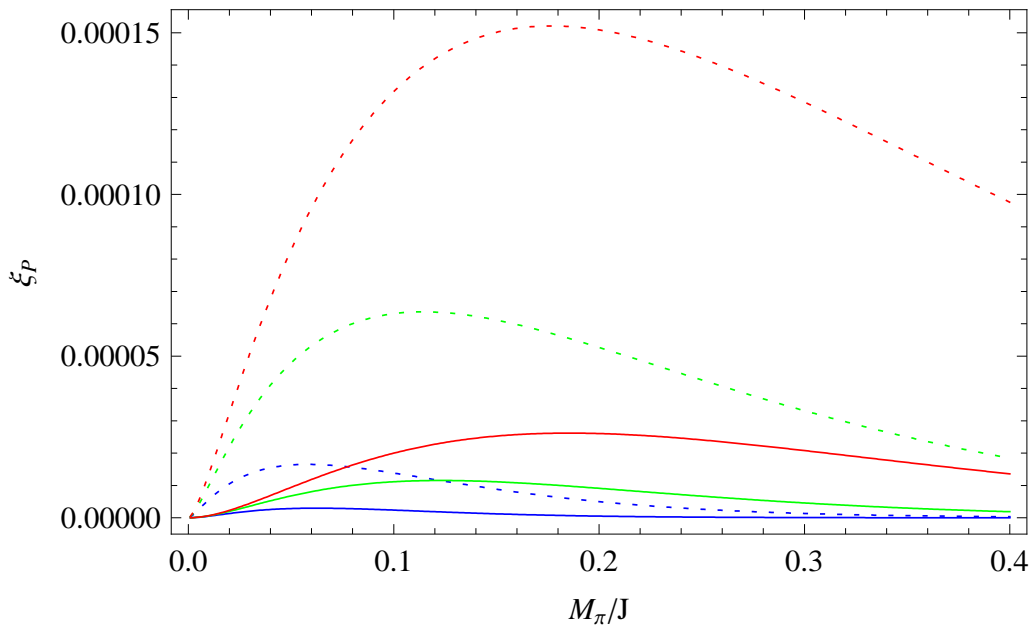}
\caption{[Color online] Quantum XY model in $d$=3+1 (continuous curves) and $d$=2+1 (dotted curves). Leading interaction corrections in
the pressure, evaluated at $T/J = \{ 0.015, 0.030, 0.045 \}$, from bottom to top in the figure.}
\label{figure9}
\end{figure}

In order to compare the strength of the spin-wave interaction in the two models, we choose the temperatures $T/J = \{ 0.015, 0.030, 0.045
\}$ in either case. We consider the leading (two-loop) interaction correction in the pressure and measure the interaction strength by
$\xi_P(T,H_s)$ defined in Eq.~(\ref{intRatioP}). In Fig.~\ref{figure9} we plot this quantity as a function of the external field. The
spin-wave interaction in the lower-dimensional case is larger, but one should keep in mind that -- with respect to the respective
transition temperatures ($T_{KT}$ versus $T_c$) -- the temperature is also more elevated in the (2+1)-dimensional model.

While the ratios $\sigma = M_{\pi}/2 \pi T$ and $\tau = T/ M_{\pi}$ can take any value in three spatial dimensions -- as long as $T$ and
$M_{\pi}$ ($H_s$) are small compared to the underlying scale $J$ -- restrictions are imposed in two spatial dimensions due to the
Mermin-Wagner theorem \citep{MW66}. In particular, as we have argued in Ref.~\citep{Hof14a}, the external field $H_s$ cannot be switched
off completely, because one would enter a domain where the present effective expansion no longer applies. Still, the subtle effects in the
order parameter and the susceptibility described above take place in a parameter regime where our formulas very well apply to the $d$=2+1
quantum XY model.

\section{Conclusions}
\label{Summary}

We have studied the low-temperature properties of the $d$=3+1 quantum XY model in the regime where the physics is dominated by spin waves.
To the best of our knowledge, all previous analyses were restricted to one-loop order in the low-temperature expansion. In particular, the
manifestation of the spin-wave interaction in thermodynamic quantities has not been addressed before.

Here we have presented a systematic effective field theory analysis of the $d$=3+1 quantum XY model at low temperatures in weak staggered
(or magnetic) fields up to three-loop order. The basic quantity from which the thermodynamic properties of the system can be derived is
the free energy density: the free Bose gas term in the low-temperature expansion is of order $T^4$, while two- and three-loop corrections
yield interaction contributions of order $T^6$ and $T^8$, respectively.

The $T^6$-correction only involves the leading-order effective constants $F$ and $\Sigma_s$. On the other hand, at order $T^8$ in the free
energy density, next-to-leading order effective constants show up that are a priori unknown. We have estimated their numerical values and
concluded that these effects of order $T^8$ are small.

The spin-wave interaction in the pressure is repulsive at low temperatures. Only in very weak staggered fields $H_s$ it may become
attractive -- these tiny effects depend on the numerical values of NLO effective constants whose sign is not determined by the symmetries
of the $d$=3+1 quantum XY model. Remarkably, in the staggered magnetization (susceptibility) the sign of the temperature-dependent
interaction contribution in weak fields becomes positive (negative). These somehow counterintuitive findings also occur in the $d$=2+1
quantum XY model, where the spin-wave interaction manifests itself in a qualitatively analogous way.

Lattice anisotropies only start showing up in the subleading Lagrangian ${\cal L}^4_{eff}$ that hence depends on additional effective
constants. These next-to-leading order effective constants slightly modify the coefficients of the low-temperature expansion beyond the
free Bose gas term -- but only if one aims at three-loop accuracy. These constants, however, do not alter the structure of the
temperature powers. This all justifies the use of a (pseudo-)Lorentz-invariant framework.

There are materials that are believed to behave like $d$=3+1 quantum XY ferromagnets \citep{WLH67,AJHC76,AJHR77,BAJC78,BJCJ81,CCBGF85}.
Unfortunately, to experimentally detect the subtle two- and three-loop effects presented here, appears to be out of question. The behavior
of any real material is more complicated than the simple quantum XY model system. Still, the effective field theory predictions could be
verified by simulating the "clean" $d$=3+1 quantum XY Hamiltonian.

\section*{Acknowledgments}
The author would like to thank J.\ Engels, E.\ E.\ Jenkins, H.\ Leutwyler, A.\ V.\ Manohar, J.\ Oitmaa, E.\ Vicari and U.-J.\ Wiese for
correspondence.

\begin{appendix}

\section{Evaluation of Partition Function Diagrams}
\label{appendixA}

General aspects of the perturbative evaluation of the partition function within effective field theory have been discussed before (see,
e.g., section 2 of Ref.~\citep{Hof10}, or appendix A of Ref.~\citep{Hof11a}). In the present appendix we focus on $d$=3+1
(pseudo-)Lorentz-invariant systems with a spontaneously broken symmetry O($N$) $\to$ O($N$-1). Partial results have been presented in
Ref.~\citep{Hof99b}.

The basic object is the thermal Goldstone boson propagator $G(x)$,
\begin{equation}
\label{ThermalPropagator}
G(x) = \sum_{n = - \infty}^{\infty} \Delta({\vec x}, x_4 + n \beta) \, , \qquad \beta = \frac{1}{T} \, ,
\end{equation}
where $\Delta(x)$ is the Euclidean Goldstone boson propagator at zero temperature. Using dimensional regularization, it can be represented
as 
\begin{equation}
\label{regprop}
\Delta (x) = (2 \pi)^{-d} \int {\mbox{d}}^d p e^{ipx} (M^2 + p^2)^{-1}
= {\int}_{0}^{\infty} \mbox{d} \rho (4 \pi \rho)^{-d/2} e^{- \rho M^2 - x^2/{4 \rho}} \, .
\end{equation}

Let us first list the contributions from the various graphs (see Fig.~\ref{figure1}) that are relevant for the free energy density up to
order $p^8$:
\begin{equation}
z_2 = - F^2 M^2 \, .
\end{equation}

\begin{equation}
z_{4A} = - \mbox{$ \frac{1}{2}$} (N-1) \, G_0 \, .
\end{equation}

\begin{equation}
z_{4B} = - (k_2 + k_3) \, M^4 \, .
\end{equation}

\begin{equation}
z_{6A} = \mbox{$ \frac{1}{8}$}(N-1) (N-3) \frac{M^2}{F^2} \, {(G_1})^2 \, .
\end{equation}

\begin{equation}
z_{6B} = (N-1) (k_2 - k_1) \frac{M^4}{F^2} \, G_1 \, .
\end{equation}

\begin{equation}
z_{6C} = {\hat c}_1 \, M^6 \, .
\end{equation}

\begin{equation}
z_{8A} = \mbox{$ \frac{1}{16}$} (N-1) (N+1) (N-5) \frac{M^2}{F^4} \, {(G_1)}^3 \, .
\end{equation}

\begin{equation}
z_{8B} = - \mbox{$ \frac{1}{4}$} (N-1) (N-3) \frac{M^2}{F^4} \, {(G_1)}^3
- \mbox{$ \frac{1}{16}$} (N-1) {(N-3)}^2 \frac{M^4}{F^4} \, {(G_1)}^2 G_2 \, .
\end{equation}

\begin{eqnarray}
\label{z8C}
z_{8C} & = & \mbox{$ \frac{1}{6}$} N (N-1) \frac{M^2}{F^4} \, {(G_1)}^3 
+ \mbox{$ \frac{1}{48}$} (N-1) (N-3) \frac{M^4}{F^4} \, J_1 \nonumber \\
& & - \mbox{$ \frac{1}{4}$} (N-1) (N-2) \frac{1}{F^4} \, J_2 \, .
\end{eqnarray}

\begin{equation}
z_{8D} = - (N-1) (2 e_1 + N e_2) \frac{1}{F^4} {(G_{\mu \nu})}^2
- (N-1) \Big[ e_1 (N-1) + e_2 - \mbox{$ \frac{1}{2}$} k_1 (N-3) \Big] \frac{M^4}{F^4} \, {(G_1)}^2 \, .
\end{equation}

\begin{equation}
z_{8E} = - \mbox{$ \frac{1}{2}$} (N -1) \Big[ (N-5) k_1 + 2k_2 \Big] \frac{M^4}{F^4} {(G_1)}^2
- \mbox{$ \frac{1}{2}$} (N-1) (N-3) (k_2 - k_1) \frac{M^6}{F^4} G_1 G_2 \, .
\end{equation}

\begin{equation}
z_{8F} = - 2(N -1) k_1 (k_2 - k_1) \frac{M^6}{F^4} G_1
- (N -1) {(k_2 - k_1)}^2 \frac{M^8}{F^4} G_2 \, .
\end{equation}

\begin{equation}
z_{8G} = (N -1) {\hat c}_0 \frac{M^6}{F^2} G_1 \, .
\end{equation}

\begin{equation}
z_{8H} = {\hat d}_0 M^8 \, .
\end{equation}
Note that these expressions involve the bare Goldstone boson mass $M$ that can be translated into the external staggered field $H_s$ by 
\begin{equation}
M^2 = \frac{{\Sigma}_s H_s}{F^2} \, .
\end{equation}
The thermodynamics of the $d$=3+1 quantum XY model is contained in the functions $G_0, G_1, G_2, G_{\mu \nu}, J_1, J_2$: they all depend in
a nontrivial way on the dimensionless ratio $M/T$. The quantity $G_1$ is the thermal propagator at the origin,
\begin{equation}
G_1 \equiv G(x) |_{x=0} \, ,
\end{equation}
while $G_2$ denotes the integral over the torus ${\cal T} = {\cal R}^{d_s} \times S^1$ (where $S^1$ is the circle defined by
$- \beta / 2 \leq x_4 \leq \beta / 2$, and $d_s$ is the spatial dimension), reading
\begin{equation}
\label{Torus}
G_2 \, = \, {\int}_{\!\!\! {\cal T}} {\mbox{d}}^d x \, \Big\{ G(x) \Big\}^2 \, .
\end{equation}
This quantity corresponds to the derivative of the thermal propagator at the origin with respect to the mass squared,
\begin{equation}
G_2 = - \frac{\mbox{d} G_1}{\mbox{d} M^2} \, .
\end{equation}
Then, $G_{\mu \nu}$ is the second derivative of the thermal propagator at the origin,
\begin{equation}
G_{\mu \nu} = \partial_{\mu} \partial_{\nu} G(x) |_{x=0} \, ,
\end{equation}
while $J_1$ and $J_2$ are the loop integrals
\begin{eqnarray}
\label{FunctionsJ}
J_1 & = & {\int}_{\!\!\!T} \! {\mbox{d}}^dx \; {\Big\{ G(x)\Big\}}^4 \, , \nonumber\\
J_2 & = & {\int}_{\!\!\!T} \! {\mbox{d}}^dx \; {\Big\{ {\partial}_{\mu} G(x) {\partial}_{\mu} G(x) \Big\}}^2 \, .
\end{eqnarray}

In order to eventually remove the dimensional regularization parameter $d$ in the above expressions, we decompose the thermal propagator
-- and all quantities obtained from there -- into a temperature-independent and a temperature-dependent piece according to\footnote{The
procedure was first described in Ref.~\citep{GL89}.}
\begin{equation}
\label{decoGDelta}
G(x) = \Delta (x) + {\overline G}(x) .
\end{equation}
Especially, at the origin $x$=0, we have
\begin{eqnarray}
\label{decompOrigen}
G_0 & = & - \frac{4}{d} \, M^4 \lambda + g_0 \, , \nonumber\\
G_1 & = & 2 M^2 \lambda + g_1 \, , \nonumber\\
G_2 & = & (2-d) \lambda + g_2 \, , \nonumber\\
G_{\mu \nu} & = & 2 M^4 \delta_{\mu \nu } \frac{\lambda}{d} + {\overline G_{\mu \nu}} \, , \nonumber\\
& & {\overline G_{\mu \nu}} = - \mbox{$ \frac{1}{2}$} \delta_{\mu \nu } g_0 + \delta^4_{\mu} \delta^4_{\nu }
(\mbox{$ \frac{1}{2}$} d g_0 + M^2 g_1) \, .
\end{eqnarray}
The kinematical functions $g_r$ refer to the $d$-dimensional noninteracting Bose gas, and are defined by
\begin{equation}
\label{FreeFunctions}
g_r(M, T) = 2 {\int}_{\!\!\! 0}^{\infty} \frac{\mbox{d} \rho}{(4 \pi \rho)^{d/2}} \, {\rho}^{r-1} \, \exp(- \rho M^2)
\sum_{n=1}^{\infty} \exp(- n^2/{4 \rho T^2}) \, .
\end{equation}
Note that the parameter $\lambda$ in (\ref{decompOrigen}) is divergent in the limit $d \to 4$,
\begin{eqnarray}
\label{lambda}
\lambda & = & \mbox{$ \frac{1}{2}$} \, (4 \pi)^{-d/2} \, \Gamma(1-{\mbox{$ \frac{1}{2}$}}d) M^{d-4} \nonumber\\
& = & \frac{M^{d-4}}{16{\pi}^2} \, \Bigg[ \frac{1}{d-4} - \mbox{$ \frac{1}{2}$} \{ \ln{4{\pi}} + {\Gamma}'(1) + 1 \}
+ {\cal O}(d\!-\!4) \Bigg] \, .
\end{eqnarray}

Finally, to remove the singularities contained in the loop integrals $J_1$ and $J_2$, we define the quantities ${\bar J}_1$ and
${\bar J}_2$ as,
\begin{eqnarray}
\label{FunctionsJbarJ}
{\bar J}_1 & = & J_1 - c_1 - c_2 g_1 + 6 (d-2){\lambda} (g_1)^2 \, , \nonumber\\
{\bar J}_2 & = & J_2 - c_3 - c_4 g_1 + \mbox{$ \frac{1}{3}$} (d+6)(d-2) {\lambda} \Big({\overline G}_{\mu \nu}\Big)^2
+ \mbox{$ \frac{2}{3}$} (d-2) {\lambda} M^4 (g_1)^2 \, .
\end{eqnarray}
The explicit expressions for the temperature-independent counterterms $c_1 \dots c_4$ are listed in Ref.~\citep{GL89}.

Using the above decompositions and collecting first all terms that are independent of $T$, we end up with the free energy density at zero
temperature,
\begin{eqnarray}
\label{vacuumEnergyDensity}
z_0 & = & - F^2 M^2
+ \mbox{$ \frac{1}{2}$} (N-1) M^4 \lambda
- (k_2 + k_3) \, M^4
+ \mbox{$ \frac{1}{2}$}(N-1) (N-3) \frac{M^6}{F^2} \, {\lambda}^2 \nonumber \\
& & + 2 (N-1) (k_2 - k_1) \frac{M^6}{F^2} \, \lambda
+ {\hat c}_1 \, M^6
 + \mbox{$ \frac{1}{2}$} (N-1) (N+1) (N-5) \frac{M^8}{F^4} \, {\lambda}^3 \nonumber \\
& & - 2 (N-1) (N-3) \frac{M^8}{F^4} \, {\lambda}^3
+ \mbox{$ \frac{1}{48}$} c_1 (N-1) (N-3) \frac{M^4}{F^4}
 - \mbox{$ \frac{1}{4}$} c_3 (N-1) (N-2) \frac{1}{F^4} \nonumber \\
& & + \mbox{$ \frac{4}{3}$} N (N-1) \frac{M^8}{F^4} \, {\lambda}^3
+ \mbox{$ \frac{1}{4}$} (N-1) {(N-3)}^2 \frac{M^8}{F^4} (d-2) \, \lambda^3 \nonumber \\
& & - 4 (N-1) (2 e_1 + N e_2) \frac{\lambda^2}{d} \frac{M^8}{F^4}
- 4 (N-1) \Big[ e_1 (N-1) + e_2 - \mbox{$ \frac{1}{2}$} k_1 (N-3) \Big] \frac{M^8}{F^4} \, \lambda^2\nonumber \\
& & - 2(N -1) \Big[ (N-5) k_1 + 2k_2 \Big] \frac{M^8}{F^4} \lambda^2
- (N-1) (N-3) (k_2 - k_1) \frac{M^8}{F^4} (2-d) \lambda^2 \nonumber \\
& & - 4(N -1) k_1 (k_2 - k_1) \frac{M^8}{F^4} \lambda
- (N -1) {(k_2 - k_1)}^2 \frac{M^8}{F^4} (2-d) \lambda \nonumber \\
& & + 2(N -1) {\hat c}_0 \frac{M^8}{F^2} \lambda
+ {\hat d}_0 M^8 + {\cal O}(p^{10}) \, .
\end{eqnarray}
Note that the leading contribution of order $p^2$ (the term $- F^2 M^2$) is finite, while all other terms are divergent as they contain
$\lambda$ as well as unrenormalized (infinite) subleading effective constants and the counterterms $c_1$ and $c_3$. However, these
divergences can be "annihilated" order-by-order in the effective expansion. For instance, at next-to-leading order $p^4$, the pole in
$\lambda$ is removed by renormalizing the combination $k_2 + k_3$ of effective constants from ${\cal L}^4_{eff}$. Then, the effective
constants ${\hat c}_0,{\hat c}_1$ and ${\hat d}_0$ that originate from ${\cal L}^6_{eff}$ and ${\cal L}^8_{eff}$, absorb further infinities.

Next we consider all terms linear in the kinematical functions $g_r$. Remarkably, all these contributions can be merged into a single
kinematical function -- namely $g_0$ -- by expressing $g_0$ through the renormalized Goldstone boson mass $M_{\pi}$,
\begin{equation}
\label{renMass}
M_{\pi}^2 = \frac{{\Sigma}_s H_s}{F^2} + \Big[ 2 (k_2 - k_1) + (N-3) \, \lambda \Big] \frac{({\Sigma}_s H_s)^2}{F^6}
+ c \frac{({\Sigma}_s H_s)^3}{F^{10}} + {\cal O}(H_s^4) \, ,
\end{equation}
rather than through the leading-order Goldstone boson mass $M$,
\begin{equation}
M^2 = \frac{\Sigma_s H_s}{F^2} \, . \nonumber 
\end{equation}
In the course of this renormalization process, one Taylor expands $g_0$ according to
\begin{eqnarray}
g_0(M_{\pi},T) & = & g_0(M,T) - \Big\{ \varepsilon_1 + \varepsilon_2 + {\cal O}(M^8) \Big\} \, g_1(M,T) \nonumber \\
& & + \mbox{$ \frac{1}{2}$} \Big\{ \varepsilon^2_1 + {\cal O}(M^{10}) \Big\} \, g_2(M,T) \, , \nonumber \\
\varepsilon_1 & = & \Big[ 2 (k_2 - k_1) + (N-3) \, \lambda \Big] \frac{M^4}{F^2} \, , \qquad \varepsilon_2 = c \frac{M^6}{F^4} \, .
\end{eqnarray}
Note that the relation
\begin{equation}
g_{r+1} = - \frac{\mbox{d} g_r}{\mbox{d} M^2}
\end{equation}
has been used. The constant $c$ in Eq.~(\ref{renMass}) corresponds to a rather lengthy expression containing terms involving $\lambda$,
subleading effective constants, as well as the counterterms $c_2$ and $c_4$ (see Eq.~(\ref{FunctionsJbarJ})). While the explicit
expression is not needed, we point out that the constant $c$ -- or $\varepsilon_2$ -- is finite. As far as $\varepsilon_1$ is concerned,
there are only two terms: the pole in $\lambda$ can be absorbed by the combination $k_2 - k_1$ of next-to-leading order (NLO) effective
constants. In the present case ($N$=2), a renormalized effective constant $\overline k$ can be defined as
\begin{equation}
\label{renormk1k2}
{\overline k} = 2 (k_2 - k_1) - \, \lambda \, ,
\end{equation}
such that the renormalized Goldstone boson mass $M_{\pi}$ takes the form
\begin{equation}
M_{\pi}^2 = \frac{{\Sigma}_s H_s}{F^2} + {\overline k} \frac{({\Sigma}_s H_s)^2}{F^6}
+ c \frac{({\Sigma}_s H_s)^3}{F^{10}} + {\cal O}(H_s^4) \qquad (N=2) \, .
\end{equation}
After these manipulations, the final result for the contributions linear in the kinematical functions simply is
\begin{equation}
z^{[1]} = - \mbox{$ \frac{1}{2}$} (N-1) g_0(M_{\pi},T) \, .
\end{equation}

Collecting all terms that are quadratic in the kinematical functions, we obtain
\begin{eqnarray}
\label{gquadratic}
z^{[2]} & = & \mbox{$ \frac{1}{8}$}(N-1) (N-3) \frac{M^2}{F^2} {(g_1)}^2 + C_1 (N-1) \frac{M^4}{F^4} {(g_1)}^2 \nonumber\\
& & + C_2 (N-1) \frac{1}{F^4} {(g_0)}^2 + C_2 (N-1) \frac{M^2}{F^4} g_0 g_1 \, ,
\end{eqnarray}
with coefficients $C_1, C_2$ given by
\begin{eqnarray}
C_1 & = & \mbox{$ \frac{1}{2}$} {(N-1)}^2 \lambda + \frac{1}{768 \pi^2} (3 N^2 + 32 N - 67) \nonumber\\
& & - (N+1) (e_1 + e_2) + k_1 - k_2 \, \nonumber\\
C_2 & = & 5 (N-2) \lambda + \frac{3}{16 \pi^2} (N-2) - 3 (2 e_1 + N e_2) \, .
\end{eqnarray}
In the kinematical functions we have again replaced the bare mass by the renormalized mass: $g_r(M,T) \to g_r(M_{\pi},T)\, , \ r=0,1$. Note
that in the present case ($N$=2) the dependence on the singular quantity $\lambda$ drops out in $C_2$ -- one concludes that the sum
$e_1 + e_2$ of NLO effective constants is finite. On the other hand, the pole in $\lambda$ contained in $C_1$ is absorbed by renormalizing
the combination $k_2 - k_1$, as before in Eq.~(\ref{renormk1k2}) that refers to mass renormalization. For $N$=2, we thus have 
\begin{equation}
\label{gquadraticN2}
z^{[2]} = - \frac{1}{8} \frac{M^2}{F^2} {(g_1)}^2 + {\hat C}_1 \frac{M^4}{F^4} {(g_1)}^2 + {\hat C}_2 \frac{1}{F^4} {(g_0)}^2
+ {\hat C}_2 \frac{M^2}{F^4} g_0 g_1 \qquad (N=2) \, ,
\end{equation}
with
\begin{eqnarray}
{\hat C}_1 & = & \frac{3}{256 \pi^2} - 3 (e_1 + e_2) - \mbox{$ \frac{1}{2}$} {\overline k} \, , \nonumber\\
{\hat C}_2 & = & -6 (e_1 + e_2) \, .
\end{eqnarray}
Note that in the terms proportional to $C_1$ and $C_2$ in Eq.~(\ref{gquadratic}), powers of $M^2$ can be replaced by powers of $M_{\pi}^2$:
taking into account the difference $M_{\pi}^2 - M^2$ in these expressions is beyond order $p^8$ that we aim at in the present study. The
exception is the first term in Eq.~(\ref{gquadratic}) where $M^2$ must be kept.

Finally collecting all terms cubic in the kinematical functions, along with the contributions involving the three-loop integrals
${\bar J}_1$ and ${\bar J}_2$, leads to
\begin{eqnarray}
z^{[3]} & = & \mbox{$ \frac{1}{48}$} (N-1) (N-3) (3N-7) \frac{M^2}{F^4} \, {(g_1)}^3
- \mbox{$ \frac{1}{16}$} (N-1) {(N-3)}^2 \frac{M^4}{F^4} \, {(g_1)}^2 g_2 \nonumber \\
& & + \mbox{$ \frac{1}{48}$} (N-1) (N-3) \, \frac{M^4}{F^4} \, {\bar J}_1
- \mbox{$ \frac{1}{4}$} (N-1) (N-2) \, \frac{1}{F^4} \, {\bar J}_2 \, .
\end{eqnarray}
Here it is legitimate to express everything in terms of $M_{\pi}$, rather than $M$, since the error one introduces is beyond order $p^8$.

After this lengthy exercise, we obtain the following representation for the free energy density up to order $p^8$ (and for general
$N \ge 2$):
\begin{equation}
\label{freeEnergyDensityFull}
z = z_0 - \mbox{$ \frac{1}{2}$} (N-1) g_0 - 4 \pi a \, (g_1)^2 - \pi b g + \frac{1}{F^4} I + {\cal O}(p^{10}) \, .
\end{equation}
While the function $g$ is a combination of the kinematical functions $g_0$ and $g_1$,
\begin{equation}
\label{gCombSimpl}
g = 3 g_0 \, (g_0 + M^2_{\pi} \, g_1) \, ,
\end{equation}
the quantities $a$ and $b$ involve effective constants from the next-to-leading order piece ${\cal L}^4_{eff}$,
\begin{eqnarray}
\label{ConstAB}
a & = & - \frac{(N-1)(N-3)}{32{\pi}} \frac{{\Sigma}_s H_s}{F^4} + \frac{N-1}{4{\pi}} \frac{({\Sigma}_s H_s)^2}{F^8} \,
\Bigg\{ \Big[(N+1)(e_1 + e_2) + k_2 - k_1 \Big] \nonumber \\
& & - \frac{(N-1)^2}{2} \, \lambda - \frac{3N^2 + 32N - 67}{768 {\pi}^2} \Bigg\} + {\cal O}(H_s^3) \, , \nonumber \\ 
b & = & \frac{N-1}{{\pi}F^4} \, \Bigg\{ ( 2 e_1 + N e_2 ) - \frac{5(N\!-\!2)}{3} \, \lambda - \frac{N-2}{16{\pi}^2} \Bigg\} \, .
\end{eqnarray}
Finally, the function $I$ reads
\begin{eqnarray}
\label{FunctionI}
I & = & \mbox{$ \frac{1}{48}$} (N-1)(N-3) M^4_{\pi} {\bar J}_1
- \mbox{$ \frac{1}{4}$} (N-1)(N-2) {\bar J}_2 \\
& & - \mbox{$ \frac{1}{16}$} (N-1)(N-3)^2 M^4_{\pi} (g_1)^2 g_2
+ \mbox{$ \frac{1}{48}$} (N-1)(N-3)(3N-7) M^2_{\pi} (g_1)^3 \, . \nonumber
\end{eqnarray}
One notices that the evaluation of ${\bar J}_2$ is not needed for $N$=2. Following the convention of Ref.~\citep{GL89}, we rewrite $I$ as 
\begin{equation}
\label{functIJ}
I = \frac{1}{{\pi}^2} g j \, ,
\end{equation}
which defines the dimensionless function $j$ that we have evaluated numerically (see appendix \ref{appendixB}). One then ends up with the
representation
\begin{equation}
\label{fedFinal}
z = z_0 - \mbox{$ \frac{1}{2}$} (N-1) g_0 - 4 \pi a (g_1)^2 - \pi g \, \Big[b - \frac{j}{{\pi}^3 F^4}\Big] + {\cal O}(p^{10}) \, .
\end{equation}

For the $d$=3+1 quantum XY model ($N$=2), the quantities $I, a, b$ amount to
\begin{eqnarray}
I & = & - \mbox{$ \frac{1}{48}$} M^4_{\pi} {\bar J}_1 - \mbox{$ \frac{1}{16}$} M^4_{\pi} (g_1)^2 g_2
+ \mbox{$ \frac{1}{48}$} M^2_{\pi} (g_1)^3 \, , \nonumber \\
a & = & \frac{1}{32{\pi}} \frac{{\Sigma}_s H_s}{F^4} + \frac{1}{4{\pi}} \frac{({\Sigma}_s H_s)^2}{F^8} \, \Bigg\{
\Big[3(e_1 + e_2) + k_2 - k_1 \Big] - \mbox{$ \frac{1}{2}$} \lambda - \frac{3}{256 \pi^2} \Bigg\} + {\cal O}(H_s^3) \, , \nonumber \\
b & = & \frac{2}{\pi F^4} \, (e_1 + e_2) \, .
\end{eqnarray}
Note again that the combination $e_1 + e_2$ of NLO effective constants appearing in $b$ is finite. On the other hand, the pole in
$\lambda$ showing up in $a$ can be absorbed into the combination $k_2 - k_1$ of bare NLO effective constants that get renormalized
according to Eq.~(\ref{renormk1k2}). The final expression for $a$ hence is
\begin{equation}
a = \frac{1}{32{\pi}} \frac{{\Sigma}_s H_s}{F^4} + \frac{1}{4{\pi}} \frac{({\Sigma}_s H_s)^2}{F^8} \, \Bigg\{
3(e_1 + e_2) + \mbox{$ \frac{1}{2}$} {\overline k} - \frac{3}{256 \pi^2} \Bigg\} + {\cal O}(H_s^3) \qquad (N=2) \, .
\end{equation}

\section{Evaluation of the Cateye Diagram in $d$=3+1}
\label{appendixB}

In this appendix we consider the renormalization and numerical evaluation of the cateye graph 8C of Fig.~\ref{figure1}. The relevant
contributions in $z_{8C}$, Eq.~(\ref{z8C}), are the singular three-loop integrals $J_1$ and $J_2$ that are defined in
Eq.~(\ref{FunctionsJ}). How to extract the finite and physical pieces from these divergent expressions has been described in detail in
Ref.~\citep{GL89}. The outcome is summarized in Eq.~(\ref{FunctionsJbarJ}), where the quantities ${\bar J}_1$ and ${\bar J}_1$ are finite.

In the free energy density, ${\bar J}_1$ and ${\bar J}_1$ are contained in the function $I$ that we have defined in Eq.~(\ref{FunctionI}).
In the present case of the $d$=3+1 quantum XY model ($N$=2), only the first contribution (${\bar J}_1$) is relevant. 

The explicit expression for the renormalized integral ${\bar J}_1$ takes the form (for details see Ref.~\citep{GL89})
\begin{eqnarray}
\label{J1bar}
{\bar J}_1 & = & {\int}_{\!\!\! {\cal T} \setminus {\cal S}} \! \! {\mbox{d}}^4 x \, U
+ {\int}_{\!\!\! {\cal S}} \! \! {\mbox{d}}^4 x \, V
- {\int}_{\!\!\! {\cal R} \setminus {\cal S}} \! \! {\mbox{d}}^4 x \, W \, ,
\nonumber \\
U & = & G^4 \, , \nonumber \\
V & = & {\overline G}^4 + 4 {\overline G}^3 \Delta + 6 ( {\overline G}^2 - g_1^2) \Delta^2 \, , \nonumber \\
W & = & 6 g_1^2 \Delta^2 + 4 g_1 ch(M x_4) \Delta^3 + \Delta^4 \, .
\end{eqnarray}
The quantities $G, {\overline G}, \Delta$ and $ g_1$ are defined in Eq.~(\ref{ThermalPropagator}), Eq.~(\ref{regprop}),
Eq.~(\ref{decoGDelta}) and Eq.~(\ref{FreeFunctions}). All terms in the above representation for ${\bar J}_1$ are finite and refer to the
limit $d \to 4$. Note that the case $d \to 3$ has been described in Ref.~\citep{Hof14a} which was devoted to the quantum XY model in
$d$=2+1.

The functions $G(x), {\overline G}(x)$ and $\Delta(x)$ only depend on the variables $r=|{\vec x}|$ and $t=x_4$. The above integrals are
therefore two-dimensional,
\begin{equation}
{\mbox{d}}^4 x = 4 \pi r^2 {\mbox{d}} r {\mbox{d}} t \, ,
\end{equation}
and can be evaluated straightforwardly. 

The function $G(x)$ is a modified Bessel function of the second kind. In terms of the dimensionless variables $\xi$ and $\eta$,
\begin{equation}
\xi = T |{\vec x}| \, , \qquad \eta = T x_4 \, ,
\end{equation}
we have
\begin{equation}
G(x) = \sigma \sum_{n=-\infty}^{\infty} \frac{K_1(z \, \sigma)}{z} \, , \qquad z = 2 \pi \sqrt{{(\eta + n)}^2 + \xi^2} \, .
\end{equation}

Notice that the radius of the sphere $|{\cal S}| \leq \beta/2$ (with $\beta = 1/T)$ that occurs in the representation for ${\bar J}_1$, is
arbitrary. As a consequence, the result for ${\bar J}_1$ must be independent thereof. This provides us with a very useful consistency
check regarding the numerical evaluation of the above integrals. By choosing different sizes of the sphere, we have checked that the final
result for ${\bar J}_1$ indeed is independent of $|{\cal S}|$.

\begin{figure}[t]
\begin{center}
\includegraphics[width=14cm]{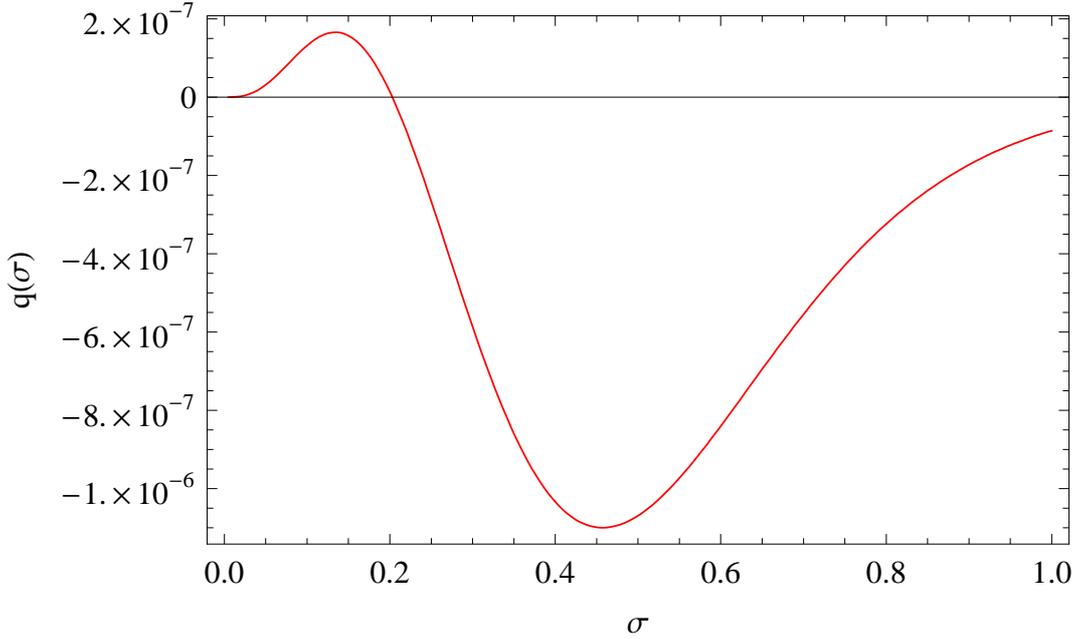}
\end{center}
\caption{The three-loop integral $q(\sigma)$ as a function of the parameter $\sigma = M_{\pi}/2\pi T$.}
\label{figureC}
\end{figure}

As described in the previous appendix (and following Ref.~\citep{GL89}), the renormalized integral ${\bar J}_1$ is contained in the
function $I$ ($N$=2),
\begin{equation}
I = - \mbox{$ \frac{1}{48}$} M^4_{\pi} {\bar J}_1 - \mbox{$ \frac{1}{16}$} M^4_{\pi} (g_1)^2 g_2
+ \mbox{$ \frac{1}{48}$} M^2_{\pi} (g_1)^3 \, , \nonumber
\end{equation}
that is rewritten by defining the dimensionless function $j=j(\sigma) \, , \ \sigma = M_{\pi}/2 \pi T$, as
\begin{equation}
I = \frac{1}{{\pi}^2} g j \, . \nonumber
\end{equation}
This function $j$ then appears in the free energy density and in all thermodynamic observables derived from there. A graph for the
function $q(\sigma)$,
\begin{equation}
T^8 q(\sigma) \equiv - \mbox{$ \frac{1}{48}$} M^4_{\pi} {\bar J}_1 \, ,
\end{equation}
is depicted in Fig.~\ref{figureC}.

\section{Estimation of NLO Effective Constants}
\label{appendixC}

The leading-order effective Lagrangian involves $F^2$ and two derivatives, while ${\cal L}^4_{eff}$ involves the NLO constants $e_i$ and
four derivatives. The derivatives correspond to powers of momenta, where the momenta are small compared to a given underlying scale
$\Lambda$. We thus have
\begin{eqnarray}
{\cal L}^2_{eff} & \propto & \frac{F^2}{2} \, p^2 = \frac{F^2}{2} \, \Lambda^2 \Big( \frac{p^2}{\Lambda^2} \Big) \, , \nonumber \\
{\cal L}^4_{eff} & \propto & e_i \, p^4 = e_i \, \Lambda^4 \Big( \frac{p^4}{\Lambda^4} \Big) \, .
\end{eqnarray}
With respect to ${\cal L}^2_{eff}$, contributions from ${\cal L}^4_{eff}$ are suppressed by $p^2/\Lambda^2$, such that we obtain
\begin{equation}
\frac{F^2}{2} \, \Lambda^2 \approx e_i \, \Lambda^4 \, ,
\end{equation}
or
\begin{equation}
\label{estimatec4}
e_i \approx \frac{F^2}{2 \Lambda^2} \, .
\end{equation}
The obvious question is which underlying scale $\Lambda$ we should choose. In analogy to quantum chromodynamics where this non-Goldstone
boson scale can be identified with the mass of the $\rho$-resonance, let us consider the ferrimagnet. The spectrum of this condensed
matter system is characterized by both acoustic and optical spin-wave excitations. While the former are Goldstone bosons, the latter are
not and are characterized by an energy gap -- a typical value is $\Delta E \approx 10 J$ \citep{Bro63}. As a typical non-Goldstone boson
scale in ferrimagnets one may thus choose $\Lambda = \Delta E \approx 10 J$. Although there are no optical spin-wave branches in the
quantum XY model, by analogy, we may still choose the representative scale as $\Lambda = 10 J$.

Now in order to estimate the value of the effective constant $F$ that also appears in Eq.~(\ref{estimatec4}), we invoke the critical
temperature $T_c$ where the order parameter drops to zero. The low-temperature expansion of the order parameter, Eq.~(\ref{OPTauN2}), at
leading order and in the absence of an external field, reduces to
\begin{equation}
\Sigma_s(T) = \Sigma_s \Big( 1- \frac{1}{24 F^2} T^2 \Big) \, .
\end{equation}
The condition $\Sigma_s(T) = 0$ then leads to
\begin{equation}
\label{estimateFJ}
\frac{F^2}{{T_c}^2} = \frac{1}{24} \, ,
\end{equation}
or
\begin{equation}
F = 0.20 \, T_c \, .
\end{equation}
For the simple cubic lattice we have $T_c \approx 2.02 J$ \citep{BEL70}, such that
\begin{equation}
F \approx 0.41 \, J \, .
\end{equation}
Hence for the NLO effective constants $e_i$ we obtain the estimate\footnote{It should be noted that we estimate the magnitude of these
NLO effective constants -- their signs remain inconclusive.}
\begin{equation}
e_i \approx \frac{F^2}{2 \Lambda^2} \approx 0.001 \, .
\end{equation}
In contrast to $e_1$ and $e_2$, the combination $\overline k$ of NLO effective constants requires logarithmic renormalization. However our
convention Eq.~(\ref{renormk1k2}) also leads to
\begin{equation}
{\overline k} \approx \frac{F^2}{2 \Lambda^2} \approx 0.001 \, .
\end{equation}
Finally we point out that the estimated order of magnitude of the NLO effective constants is consistent with the general power counting
rules derived in Ref.~\citep{GJMM16} that imply
\begin{equation}
\Lambda = 4 \pi F \, , \qquad e_i \approx \frac{F^2}{2 \Lambda^2} \approx \frac{1}{32 \pi^2} \approx 0.003 \, .
\end{equation}

\end{appendix}

\end{document}